\documentclass[12pt,preprint]{aastex}  
  
\newcommand{\msun}{M_{\odot}} 
 
\shorttitle{The Asiago-ESO/RASS QSO Survey III}   
\shortauthors{Grazian et al.}   
   
\begin{document}   
   
   
\title{The Asiago-ESO/RASS QSO Survey \\   
III. Clustering analysis and its theoretical interpretation\footnote{Based   
on observations collected at the European Southern Observatory, Chile   
(ESO P66.A-0277 and ESO P67.A-0537), with the Arizona Steward   
Observatory and with National Telescope Galileo (TNG) during AO3   
period.}}   
   
\author{Andrea Grazian}   
\affil{INAF-Osservatorio Astronomico di Roma, 
via Frascati 33, I--00040 Monte Porzio Catone, Italy 
\and 
Dipartimento di Astronomia, Universit\`{a} di Padova,   
vicolo dell'Osservatorio 2, I--35122 Padova,   
Italy}   
\email{grazian@mporzio.astro.it, grazian@pd.astro.it}   
   
\author{Mattia Negrello}   
\affil{S.I.S.S.A., via Beirut 4, I--34014 Trieste, Italy}   
\email{negrello@sissa.it}   
   
\author{Lauro Moscardini}   
\affil{Dipartimento di Astronomia, Universit\`{a} di Bologna,    
via Ranzani 1, I--40127 Bologna, Italy}   
\email{moscardini@bo.astro.it}   
   
\author{Stefano Cristiani}   
\affil{INAF-Osservatorio Astronomico di Trieste,    
	       via G.B. Tiepolo 11, I--34131 Trieste, Italy}   
\email{cristiani@ts.astro.it}   
   
\author{Martin G. Haehnelt}  
\affil{Institute of Astronomy, Madingley Road, Cambridge CB3 0HA, England}  
\email{haehnelt@ast.cam.ac.uk}  
 
\author{Sabino Matarrese}   
\affil{Dipartimento di Fisica ``G. Galilei'' and I.N.F.N., Sezione di  
Padova, Universit\`{a} di Padova, via Marzolo 8, I--35131 Padova, Italy} 
\email{matarrese@pd.infn.it}   
   
\author{Alessandro Omizzolo}   
\affil{Vatican Observatory Research Group, University of Arizona,   
	       Tucson AZ 85721, USA 
\and 
Dipartimento di Astronomia, Universit\`{a} di Padova,    
vicolo dell'Osservatorio 2, I--35122 Padova,   
Italy}   
\email{aomizzolo@specola.va, omizzolo@pd.astro.it}   
   
\author{Eros Vanzella}   
\affil{European Southern Observatory, Karl-Schwarzschild Str. 2,   
 D--85748 Garching, Germany   
\and   
Dipartimento di Astronomia, Universit\`{a} di Padova,    
vicolo dell'Osservatorio 2, I--35122 Padova,   
Italy}   
\email{evanzell@eso.org, vanzella@pd.astro.it}   
   
   
   
\begin{abstract}   
This is the third paper of a series describing the Asiago-ESO/RASS QSO 
survey (AERQS), a project aimed at the construction of an all-sky 
statistically well-defined sample of relatively bright QSOs ($B\le 
15$) at $z\le 0.3$.  We present here the clustering analysis of the 
full spectroscopically identified database (392 AGN).  The clustering 
signal at $0.02<z<0.22$ is detected at a $3-4\sigma$ level and its 
amplitude is measured to be $r_0=8.6\pm 2.0 h^{-1}$ Mpc (in a $\Lambda 
CDM$ model).  The comparison with other classes of objects shows that 
low-redshift QSOs are clustered in a similar way to Radio Galaxies, 
EROs and early-type galaxies in general, although with a marginally 
smaller amplitude.  The comparison with recent results from the 2QZ 
shows that the correlation function of QSOs is constant in redshift or 
marginally increasing toward low redshift.  We discuss this behavior 
with physically motivated models, deriving interesting constraints on 
the typical mass of the dark matter halos hosting QSOs, $M_{\rm 
DMH}\sim 10^{12.7} h^{-1} \msun$ ($10^{12.0}-10^{13.5}h^{-1} M_{\odot}$ at 
1$\sigma$ confidence level). Finally, we use the clustering data 
to infer the physical properties of local AGN, obtaining $M_{\rm 
BH}\sim 2\cdot 10^8 h^{-1} \msun$ 
($10^7-3\cdot 10^9 h^{-1} \msun$) for the mass of the active black 
holes, $\tau_{AGN}\sim 8\cdot 10^6$ yr 
($ 2\cdot 10^{6}-5\cdot 10^{7}$ yr) for their life-time and 
$\eta \sim 0.14$ for their efficiency (always for a $\Lambda CDM$ model). 
 
\end{abstract}   
   
   
   
   
\section{Introduction}   
   
The analysis of the statistical properties (luminosity function and   
clustering) of the cosmic structures is a fundamental cosmological   
tool to understand their formation and evolution. The clustering of   
QSOs and galaxies at small to intermediate scales (1-50 $h^{-1}$ Mpc)   
provides detailed information on the distribution of Dark Matter Halos   
(DMHs) that are generally thought to constitute the ``tissue'' on   
which cosmic structures form.  This means to investigate   
--indirectly-- fundamental astrophysical problems, such as the nature   
of dark matter, the growth of structures via gravitational   
instability, the primordial spectrum of density fluctuations and its   
transfer function.  The lighting up of galaxies and other luminous   
objects, such as QSOs, involves complex and non-linear physics. It   
depends on how the baryons cool within the DMHs and form stars or   
start accreting onto the central black hole (BH), ending up as the   
only directly visible peak of a much larger, invisible structure.   
The so-called bias factor, $b(r,z)$, is used to   
explain the difference between visible structures and invisible   
matter, whose gravity governs the overall evolution of   
clustering. This complex relation is summarized by the simple formula   
$\xi(r,z)=b^{2}(r,z)\xi_{\rm m}(r,z)$, where $\xi(r,z)$ and $\xi_{\rm   
m}(r,z)$ are the two-point correlation functions (TPCF) of radiating   
objects and dark matter, respectively.   
In this way the detailed analysis of the distribution of the peaks of   
visible matter can distinguish among the various models for the   
formation of structures. In particular the {\em hierarchical growth of   
structures} is naturally predicted in a cold dark matter (CDM)   
scenario, where larger objects are constantly formed from the assembly   
of smaller ones.  An alternative view of the structure formation and   
evolution, supported by both some observed properties of high-redshift   
ellipticals and EROs \citep{daddi01,daddi02} and theoretical modeling   
\citep{lynden,larson,matteucci,tantalo}, leads to the scenario of {\em   
monolithic collapse}, i.e. an earlier object formation and a following   
passive evolution.  The clustering data can be used to discuss whether the   
merging processes were important at various redshifts or the galaxy   
number tends to be conserved during the evolution. These two opposite   
models predict a significantly different redshift evolution of the bias   
factor \citep{mclm97,mclm98}.   
   
The first attempt to measure the clustering of QSOs was made by   
\citet{osmer81}. \citet{s84} was the first to detect QSO clustering on small   
scales using the \citet{VV84} catalog, a collection of inhomogeneous 
samples.  A number of authors 
\citep{is88,ac92,mf93,sb94,andreani,cs96} have used complete and 
better defined QSO samples to measure their spatial distribution. At a 
mean redshift of $z\sim 1.4$, they generally detect a clustering 
signal at a typical significance level of $\sim3-4\sigma$, 
corresponding to a correlation length, $r_0$, similar to the value 
obtained for local galaxies: $r_0\sim6 h^{-1}$ Mpc. However, there has 
been significant disagreement over the redshift evolution of QSO 
clustering, including claims for a decrease of $r_0$ with redshift 
\citep{is88}, an increase of $r_0$ with redshift \citep{lac98} and no 
change with redshift \citep{cs96}. 
Recently, \citet{2qzclust}, using more than 10,000 objects taken from  
the preliminary data release catalog of 2dF QSO Redshift Survey  
(hereafter 2QZ), measured the evolution of QSO clustering as a  
function of redshift. Assuming an Einstein-de Sitter universe  
($\Omega_{M}=1.0$ and $\Omega_{\Lambda}=0.0$), they found no  
significant evolution for $r_0$ in comoving coordinates over the  
redshift range $0.3\le z\le 2.9$, whereas for a model with  
$\Omega_{M}=0.3$ and $\Omega_{\Lambda}=0.7$ the clustering signal  
shows a marginal increase at high redshift.  Here $\Omega_{M}$ and  
$\Omega_{\Lambda}$ are the mass and cosmological constant density  
contribution, respectively, to the total density of the universe.  
   
The observed behavior of the QSO clustering can be explained within  
the linear theory and a typical bias model.  The theoretical  
interpretation of the picture drawn by 2QZ is a result of the  
combination of many ingredients and their degeneracies: the bias  
factor, the ratio between the masses of black hole and dark matter  
halo, the life-time of QSOs, the efficiency and the mass accretion  
rate.  
   
To add new insights in the modeling and interpretation, one has to   
consider the constraints from the luminosity function (LF) or/and to   
enlarge the redshift domain toward lower or higher redshifts. For   
these reasons, we have started a project, the AERQS, to find bright   
AGN in the local universe, removing present   
uncertainties about the properties of the local QSO population and   
setting the zero point for clustering evolution. For the general aims   
of the AERQS Survey and its detailed presentation [see   
\citet{GC2000,DSS2k2}].   
   
The goal of this paper is to analyze the clustering properties of a  
well defined large sample of bright QSOs at $z\le 0.3$, and provide  
key information on the following issues: {\em what is the typical mass  
of DMHs hosting AGN?  What is the typical bias factor for AGN?  What  
is the duty cycle for AGN activity?  What is the typical efficiency of  
the central engine at the various redshifts?}  
   
The plan of the paper is as follows. In \S\ 2 we describe the data  
used in the statistical analysis. The various techniques used to  
investigate the clustering properties are presented in \S\ 3, while  
\S\ 4 is devoted to a comparison with similar results obtained by   
previous surveys at low redshifts. To investigate the redshift   
evolution of the clustering, the spatial properties of QSOs in the   
local universe are compared in \S\ 5 to the recent 2QZ results at   
intermediate redshifts for QSOs and to various estimates for normal   
and peculiar galaxies.  In \S\ 6 physically motivated models are used   
to link the galactic structures at high-$z$ with the local AGN and   
galaxy population.  In \S\ 7 we discuss the clustering properties of   
QSOs in the light of these simple theoretical models.  Finally \S\ 8   
gives some concluding remarks on the clustering of QSOs.   
   
\section{The Data}   
   
It is paradoxical that in the era of 2QZ and Sloan Digital Sky Survey 
(SDSS), with thousands of faint QSOs discovered up to the highest 
redshifts, there are still relatively few bright QSOs known at low 
redshift.  One of the main reasons, as shown in previous papers 
\citep{GC2000,DSS2k2}, is the rather low surface density of low-$z$ 
and bright QSOs, of the order of few times $10^{-2}$ per 
deg$^{2}$. This corresponds to a very small number of objects also in 
the case of the 750 deg$^{2}$ of the complete 2QZ and 1000 deg$^{2}$ 
of the SDSS (during commissioning phase).  One more reason, not less 
important than the previous one, is that with the optical information 
only it is difficult to efficiently isolate bright QSOs from 
billions of stars in large areas. As a consequence, a survey based on 
different selection criteria is required. In 
\citet{GC2000,DSS2k2} we have used the X-ray emission, a key feature   
of the AGN population.   
   
\begin{figure}   
\epsscale{1.}   
\plotone{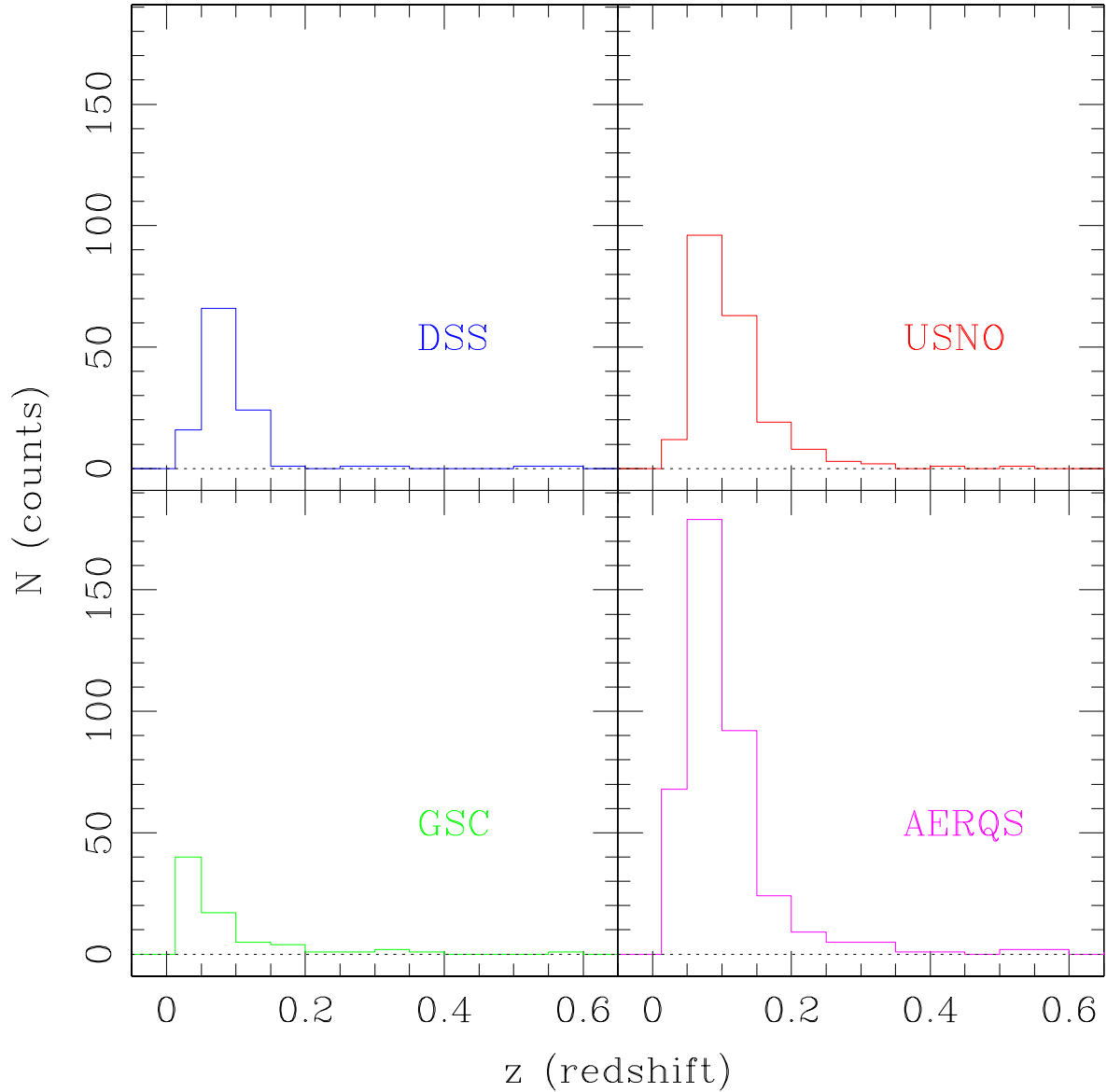}   
\caption{The redshift distributions for the three separated sub-samples   
(DSS, USNO and GSC) and for the total sample (AERQS, bottom-right  
panel). Five AGN with higher redshifts (in the range $0.6\le z\le  
2.04$) are not plotted here.  
\label{histz}}   
\end{figure}

The AERQS is divided in three sub-samples, two in the northern   
hemisphere (USNO and GSC), described in \citet{GC2000} (hereafter   
Paper I), and the DSS sample in the southern hemisphere, described in   
\citet{DSS2k2} (Paper II). After a campaign of spectroscopic   
identifications at various telescopes, we have completed the sample, 
which is made up of 392 AGN with redshifts between 0.007 and 2.043. 
The redshift distributions, shown in Fig. \ref{histz}, show a peak 
around $z\sim 0.1$ with an extended tail up to $z=0.4$. Five AGN with 
$0.6\le z\le 2.04$ are possibly objects magnified by gravitational 
lensing effects. Table \ref{aerqs} summarizes the basic properties of 
the three sub-samples.  The area covered by the AERQS Survey consists 
of $\sim 14,000$ deg$^2$ at the high Galactic latitudes ($|b_{\rm 
gal}|\ge 30^{\circ}$). The mean values for the completeness and 
efficiency are 65.7\% and 52.3\%, respectively. 
   
\begin{deluxetable}{cccccccc}   
\tablecolumns{8}   
\tablewidth{0pc}   
\tablecaption{A summary of the AERQS Survey: the three sub-samples.   
\label{aerqs}}   
\tablehead{   
\colhead{Name} & \colhead{$DEC$} & \colhead{limit magnitude} &   
\colhead{Area} &   
\colhead{$N_{AGN}$} & \colhead{Redshift} & \colhead{Completeness}}   
\startdata   
DSS & $-90\le\delta\le 0$ & $12.60\le B\le 15.13$ & 5660 & 111 &   
$0.012\le z\le 0.680$ & 0.63 \\   
USNO &   
$0\le\delta\le +90$ & $13.50\le R\le 15.40$ & 8164 & 209 & $0.034\le   
z\le 2.043$ & 0.68 \\   
GSC & $0\le\delta\le +90$ & $12.50\le   
V\le 14.50$ & 8164 & 72 & $0.007\le z\le 0.573$ & 0.63 \\   
\enddata   
\tablecomments{The reported area (in deg$^{2}$)   
is the fraction of the northern and southern hemispheres with $|b_{\rm   
gal}|\ge 30^{\circ}$ and exposure time of the ROSAT All Sky Survey   
$t_{\rm exp}\ge 300$ sec (as described in Paper I and Paper II).}   
\end{deluxetable}   
   
\section{Measuring the Clustering in the AERQS}   
   
The simplest way to analyze the clustering properties of a   
homogeneous and complete sample of QSOs is to compute the TPCF,   
$\xi(r)$, in the redshift space.  We choose to calculate $\xi(r)$ for   
two representative cosmological models: $(\Omega_{M},   
\Omega_{\Lambda})=(1.0,0.0)$ and $(0.3,0.7)$.  We will call these   
cosmological models Einstein-de Sitter (hereafter EdS) and $\Lambda$,  
respectively.  
  
To compute $\xi(r)$ we have used the minimum variance estimator   
suggested by \citet{ls93}:   
\begin{equation}   
\xi(r)=\frac{QQ(r)-2QR(r)+RR(r)}{RR(r)},   
\end{equation}   
where $QQ$, $QR$ and $RR$ are the number of QSO-QSO, QSO-random and  
random-random pairs with a separation $r\pm\Delta r$, respectively.  
Here $r$ is the comoving distance of two QSOs in the redshift space.  
We compute the TPCF in bins of $\Delta r=5 h^{-1}$ Mpc,  
where $h$ is the Hubble constant, in units of 100 km s$^{-1}$ Mpc$^{-1}$.  
The adopted values for the Hubble constant are $h=0.5$ for the EdS model,  
and $h=0.65$ for $\Lambda$. We generate 100 random samples and we  
use the mean values of $QR(r)$ and $RR(r)$ for the estimator.  
 
\begin{figure}   
\epsscale{1.00}   
\plotone{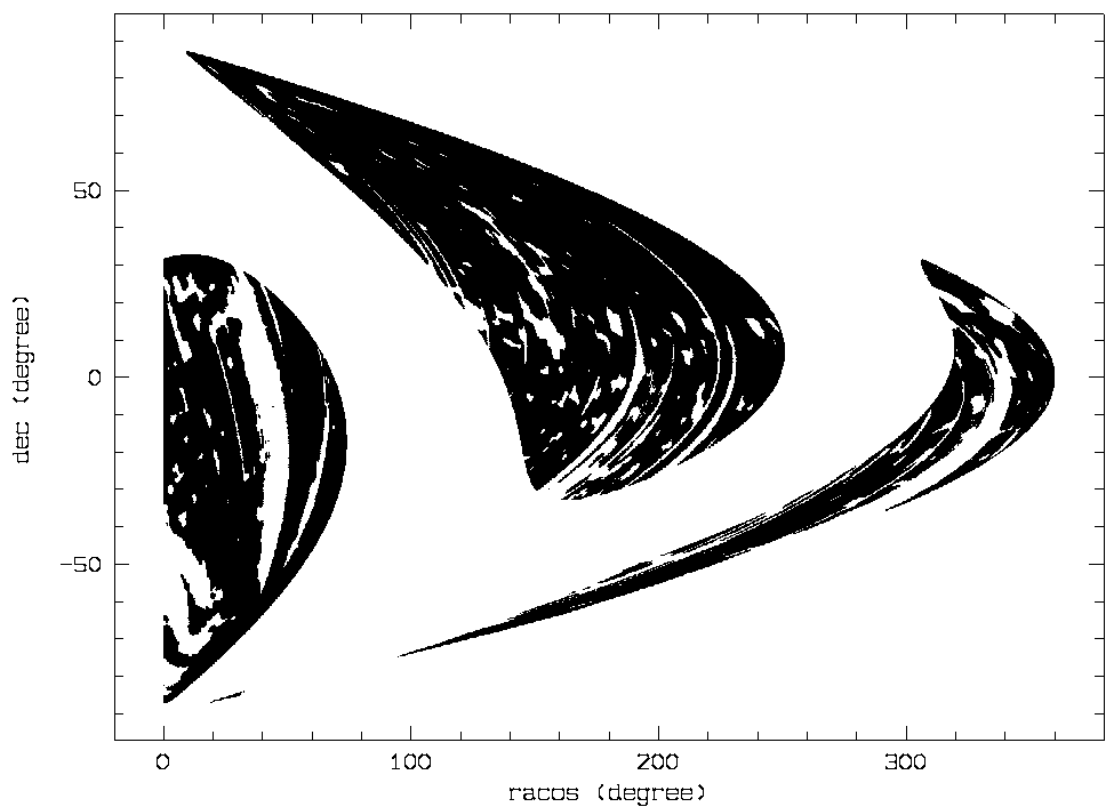}   
\caption{The black area shows the regions covered by the AERQS All Sky   
Survey after applying the selection criteria described in Paper I and   
Paper II. The projection is done here in $RA\cos(DEC)$ vs. $DEC$.   
\label{areatot}}   
\end{figure}

The correct generation of the random objects is in general the most   
critical aspect in the clustering analysis. This problem becomes   
fundamental in the case of a flux-limited sample, the AERQS.  The area 
covered by our survey is not homogeneously distributed in the sky, due   
to the selection criteria adopted and the variable Galactic   
extinction. Consequently the ``true'' apparent magnitude limit of our   
survey is variable.  Fig. \ref{areatot} shows the effective area   
covered by the AERQS survey, limited by an exposure time $t_{\rm   
exp}\ge 300$ sec in the RASS-BSC \citep{vog99} and at high Galactic 
latitudes ($|b_{\rm gal}|\ge 30^{\circ}$). 
   
The results on clustering reported in this paper are derived by   
scrambling the redshifts, and the right ascension ($RA$) and   
declination ($DEC$) coordinates for the total (AERQS) sample. The   
random $RA$ and $DEC$ are derived from Fig. \ref{areatot}, while the   
redshifts are randomly extracted from the observed (Gaussian smoothed)   
redshift distributions (Fig. \ref{histz}).   
   
In order to check the robustness of the results, we have carried out a   
more complex generation of random QSOs, which ensured the uniformity   
of the ``synthetic'' samples. The angular positions were chosen again   
randomly from the map in Fig. \ref{areatot}.  Then, for each object we   
generated random values for redshift and absolute magnitude   
reproducing the LF estimated by \citet{LFC97} and \citet{GC2000} in   
the redshift range $0.04\le z\le 2.2$. In particular for $\Phi   
(M_B,z)$ we adopt a double power-law relation evolving accordingly to   
a Luminosity Dependent Luminosity Evolution (LDLE) model:   
   
\begin{equation}   
\Phi (M_B,z)=   
{ {\Phi^\ast} \over { 10^{0.4[M_B-M_B^*(z)](\alpha + 1)} +   
10^{0.4[M_B-M_B^*(z)](\beta + 1)} } } \ , \\   
\end{equation}   
where   
\begin{equation}   
M_B^*(z) = M^*_B(z=2)-2.5 k \log [(1+z)/3]   
\end{equation}   
and   
\begin{eqnarray}   
\nonumber k = k_1 + k_2 [M_B-M_B^*(z)]e^{-{z/0.4}}&    
 {\rm if}~M_B \leq M_B^*(z) \\   
\nonumber k = k_1 & {\rm if}~M_B > M_B^*(z)\ .   
\end{eqnarray}   
   
The parameters $\alpha$ and $\beta$ correspond to the faint-end and   
bright-end slopes of the optical LF, respectively, and $M^*_B(z=2)$ is   
the magnitude of the break in the double power-law shape of the LF at   
$z=2$.  The actual values adopted in the LDLE parameterization,   
reported in Tab. \ref{ldle}, are derived by a fit to the observed LF.   
Extinction by Galactic dust is taken into account using the reddening   
$E(B-V)$ as a function of position, calculated by \citet{schlegel98}.   
   
\begin{deluxetable}{ccccccc}   
\tablecolumns{7}   
\tablewidth{0pc}   
\tablecaption{The parameters used for the LF of QSOs.   
\label{ldle}}   
\tablehead{   
\colhead{Model} & \colhead{$\Phi^\ast$} & \colhead{$M^*_B(z=2)$} &   
\colhead{$\alpha$} & \colhead{$\beta$} & \colhead{$k_1$} &   
\colhead{$k_2$}}   
\startdata   
EdS       &  9.8 & -26.3 & -1.45 & -3.76 & 3.33 & 0.37 \\   
$\Lambda$ &  5.0 & -26.7 & -1.45 & -3.76 & 3.33 & 0.30 \\   
\enddata   
\tablecomments{${\Phi^\ast}$ is in units of $10^{-7}$ mag$^{-1}$ Mpc$^{-3}$.    
}   
\end{deluxetable}   
   
\begin{deluxetable}{cccccccccc}   
\tablecolumns{9}   
\tablewidth{0pc}   
\tablecaption{A summary of the clustering properties of the AERQS Sample.   
\label{tabclust}}   
\tablehead{   
\colhead{($\Omega_{M}$,$\Omega_{\Lambda}$)} &   
\colhead{$r_0$} & \colhead{$r_{\rm low}-r_{\rm up}$} & \colhead{$\gamma$} &   
\colhead{$\bar{z}$} & \colhead{$z_{\xi}$} &   
\colhead{$\bar{\xi}(20)$} &   
\colhead{$\bar{\xi}_{\rm low}-\bar{\xi}_{\rm up}$}& bias }   
\startdata   
(1.0,0.0) &  8.49 & 6.44--10.46 & 1.58 & 0.089 & 0.063 & 0.368 &   
0.151--0.585 &  $1.75\pm 0.51$  \\   
(0.3,0.7) & 8.64 & 6.56--10.64 & 1.56 & 0.088 & 0.062 & 0.461 &   
0.224--0.698 & $1.37\pm 0.35$ \\   
\enddata   
\tablecomments{The distances are in units of $h^{-1}$ Mpc.   
The best fit value $r_0$ and its 1$\sigma$ confidence level $r_{\rm  
low}-r_{\rm up}$ are computed from the differential TPCF and the MLE  
method, assuming a fixed value for the slope $\gamma$.  The values  
$\bar{z}$ and $z_{\xi}$ are the mean redshift of the QSO sample and  
the median redshift of the observed QSO pairs inside $20 h^{-1}$ Mpc,  
respectively.  The value reported in $\bar{\xi}(20)$ is the observed  
value of the TPCF integrated over $20h^{-1}$ Mpc, with its 1$\sigma$  
confidence level, $\bar{\xi}_{\rm low}-\bar{\xi}_{\rm up}$. The bias  
factor is computed assuming the cosmological parameters described in  
Section 6.1.}  
\end{deluxetable}   
   
\begin{figure}   
\epsscale{0.80}   
\plotone{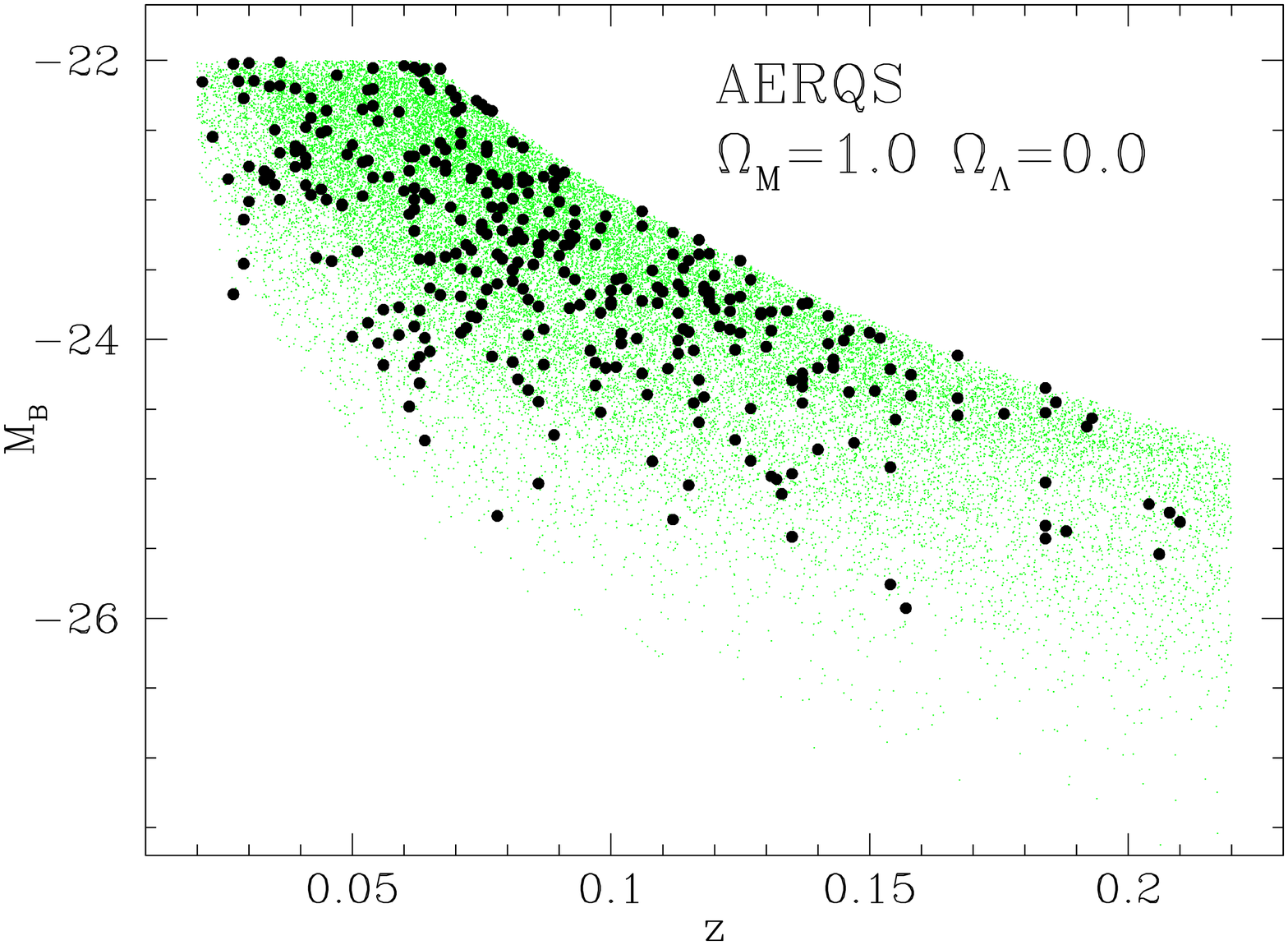}   
\caption{The redshift vs. magnitude $M_B$ distribution for the QSOs of AERQS.   
The observed AGN sample (filled circles) is compared with the random 
generated sample (small dots) in the $(z,M_B)$ space. The density of 
random points is 100 times larger than the observed one. Results are 
shown for the EdS model. 
\label{mbzflat}}   
\end{figure}

This approach, though computationally expensive, avoids biases in the   
generation of random samples of QSOs. It reproduces the   
observed LF and the distribution of redshifts and apparent magnitudes.   
Fig. \ref{mbzflat} shows the observed and random generated QSOs in the   
$(z,M_B)$ space in the case of the EdS model.   
   
The results on the clustering of QSOs derived with this particular   
approach are consistent with the ones obtained with the scrambling of   
the redshifts. In the following all the computations will be carried out   
with the latter method.   
  
\begin{figure}   
\epsscale{1.05}   
\plotone{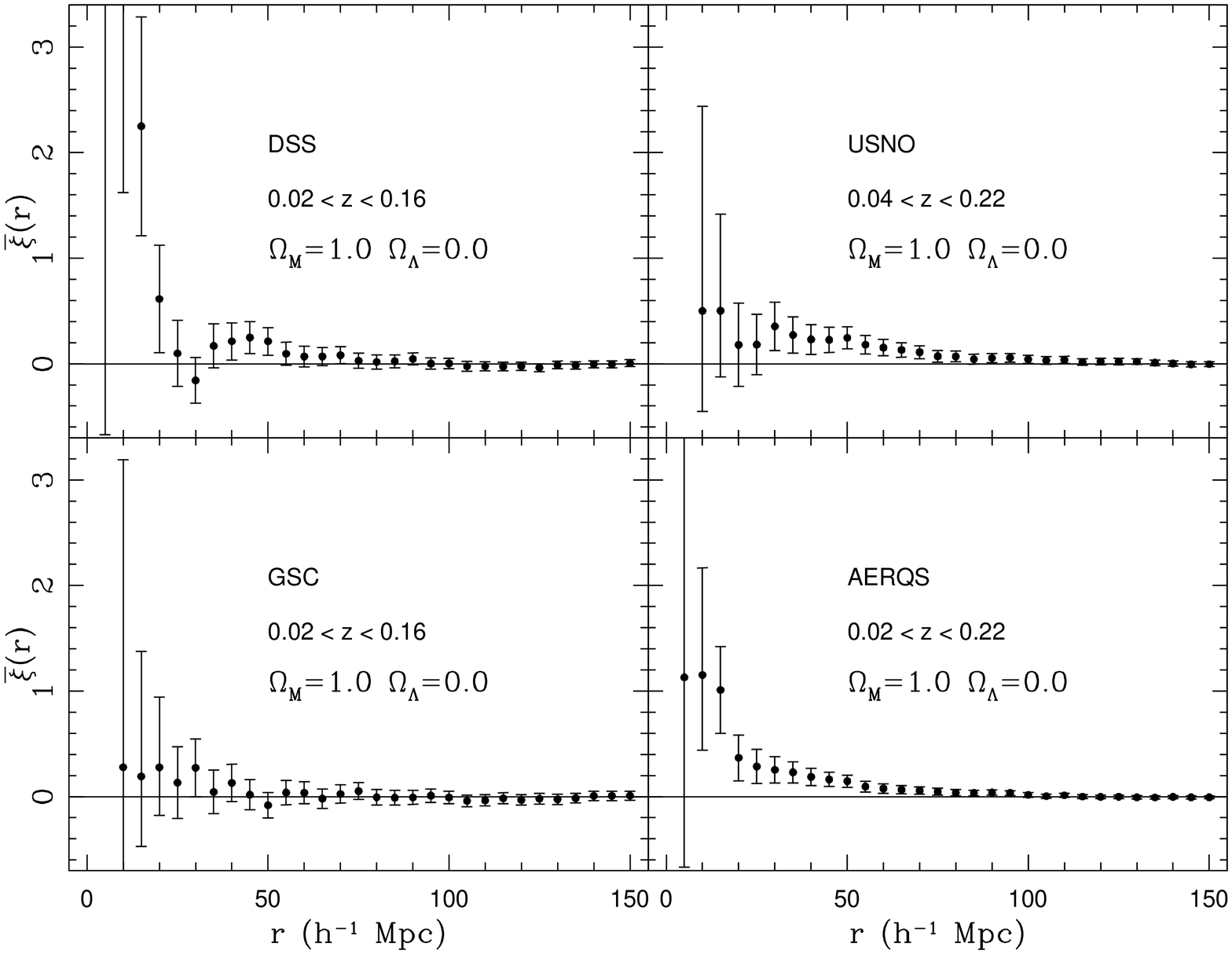}   
\caption{The integrated correlation function $\bar{\xi}(r)$, as a    
function of the sphere radius $r$, for an EdS model. Different panels 
refer to the integrated correlation function (and 1$\sigma$ error 
bars) for the DSS, GSC, USNO sub-samples, and for the AERQS total 
sample. $\bar{\xi}(r)$ is integrated over spheres 
of increasing radii. Consequently the error bars, which are shown 
only for reference, are not independent. At large scales 
($\ge 50 h^{-1} Mpc$) the integrated TPCF is consistent with zero, 
showing the absence of large-scale gradients in the data.} 
\label{totflat}   
\end{figure}

\begin{figure}   
\epsscale{1.05}   
\plotone{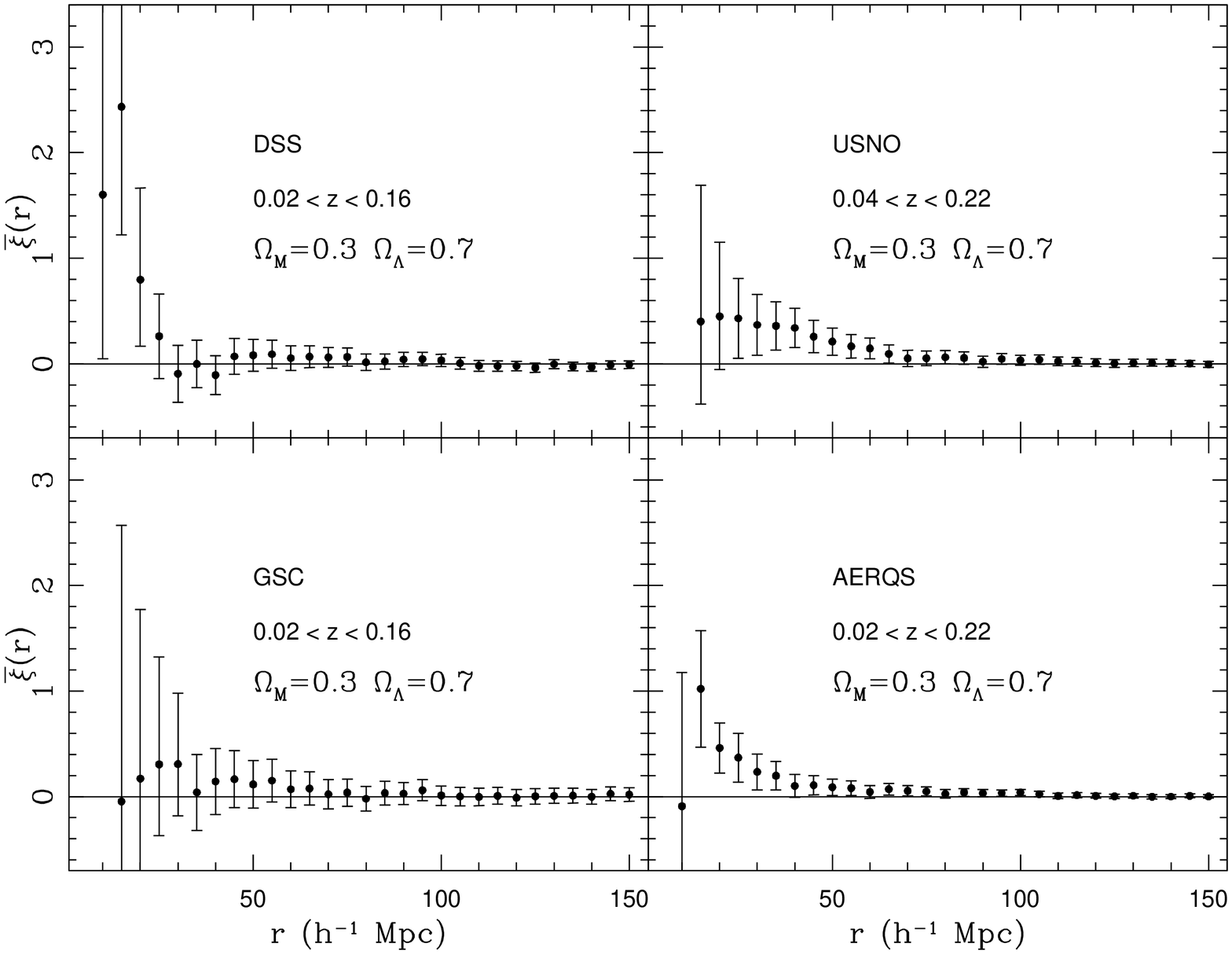} 
\caption{As Fig. \ref{totflat}, but for the $\Lambda$ model.   
\label{totlambda}}   
\end{figure}

First, we calculate the TPCF integrated over a sphere, $\bar{\xi}(r)$,   
as a function of the sphere radius $r$, for the three sub-samples   
separately (DSS, USNO, GSC) and the total sample (AERQS). Fig.  
\ref{totflat} reports the results for the EdS universe, while Fig.  
\ref{totlambda} refers to a $\Lambda$ universe.  
The error bars in Fig. \ref{totflat} and \ref{totlambda} represent the 
1$\sigma$ interval for $\bar{\xi}(r)$ and are obtained by assuming a 
Poisson distribution \citep{gehrels}. 
 
\begin{figure} 
\epsscale{1.} 
\plotone{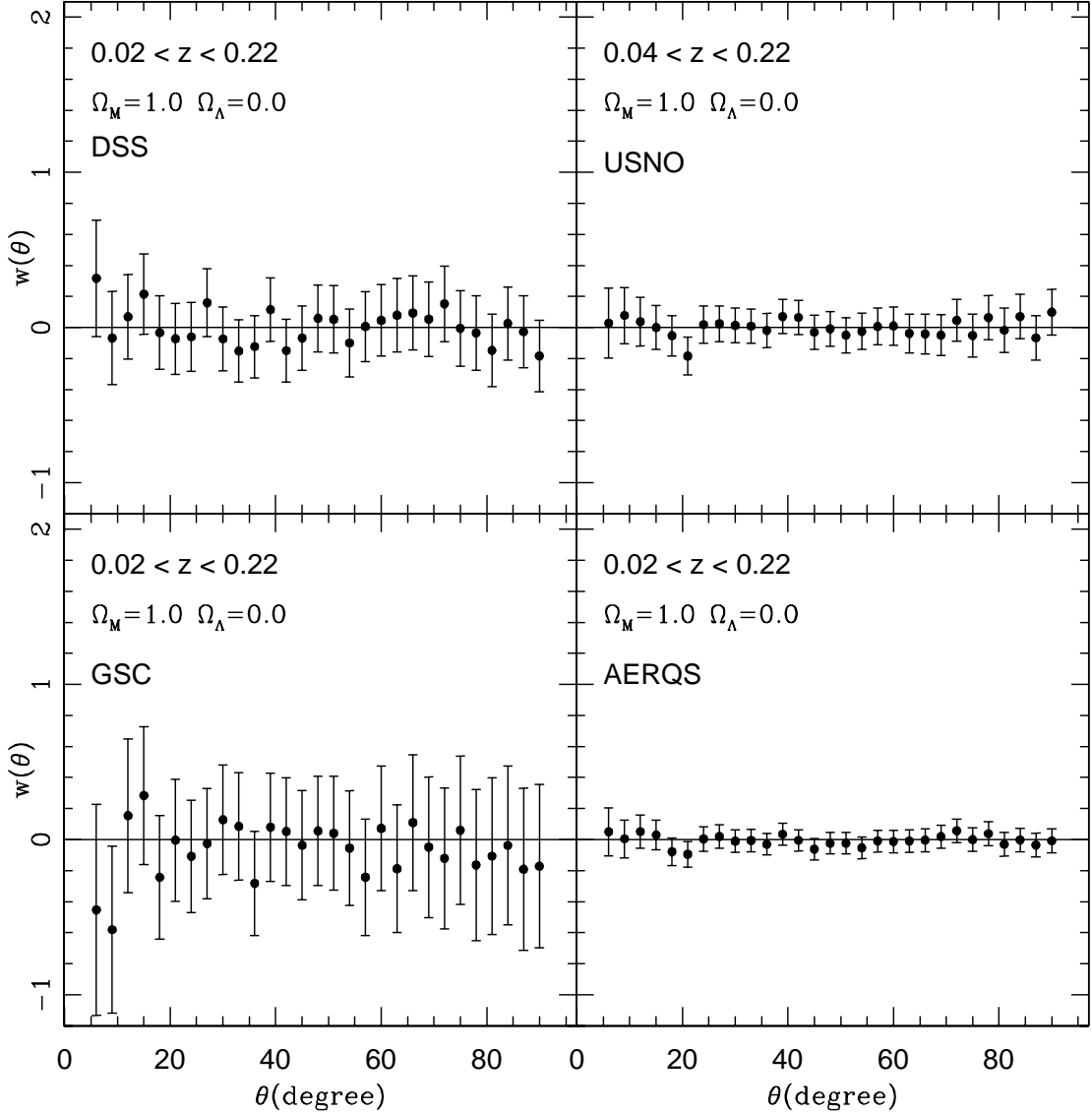} 
\caption{The differential angular TPCF binned in intervals of 3 degrees 
(corresponding to $\sim 12.6 h^{-1}$ Mpc). Different panels 
refer to the correlation function (and 1$\sigma$ error 
bars) for the DSS, GSC, USNO sub-samples, and for the AERQS total 
sample. These results do not depend on the adopted cosmological model. 
The angular TPCF at small scales is consistent with zero because 
it is diluted over 10-20 $h^{-1}$ Mpc. 
\label{angclust}} 
\end{figure}

In order to investigate the possible presence of a spurious 
clustering signal at large scales, we have computed the 
angular TPCF binned in intervals of 3 degrees (corresponding to $\sim 
14.6 h^{-1}$ Mpc comoving). Fig. \ref{angclust} shows the absence of 
any significant bias on the large scales sampled by AERQS, up to 
150 $h^{-1}$ Mpc.
  
The signal shown in Fig. \ref{totflat} and \ref{totlambda} 
for separations smaller than $15 h^{-1}$ Mpc is due to 25 and 28 QSO   
pairs for the EdS and $\Lambda$ models, respectively. For a completely   
random distribution, the expected number of pairs is 12 for EdS and 14   
for the $\Lambda$ model. Considering separations smaller than $20   
h^{-1}$ Mpc the observed pairs are 36 and 38, to be compared with 27 and 26   
random pairs expected. The clustering signal is therefore detected at a 
$3-4\sigma$ level.   
   
\begin{figure}   
\epsscale{1.00}   
\plotone{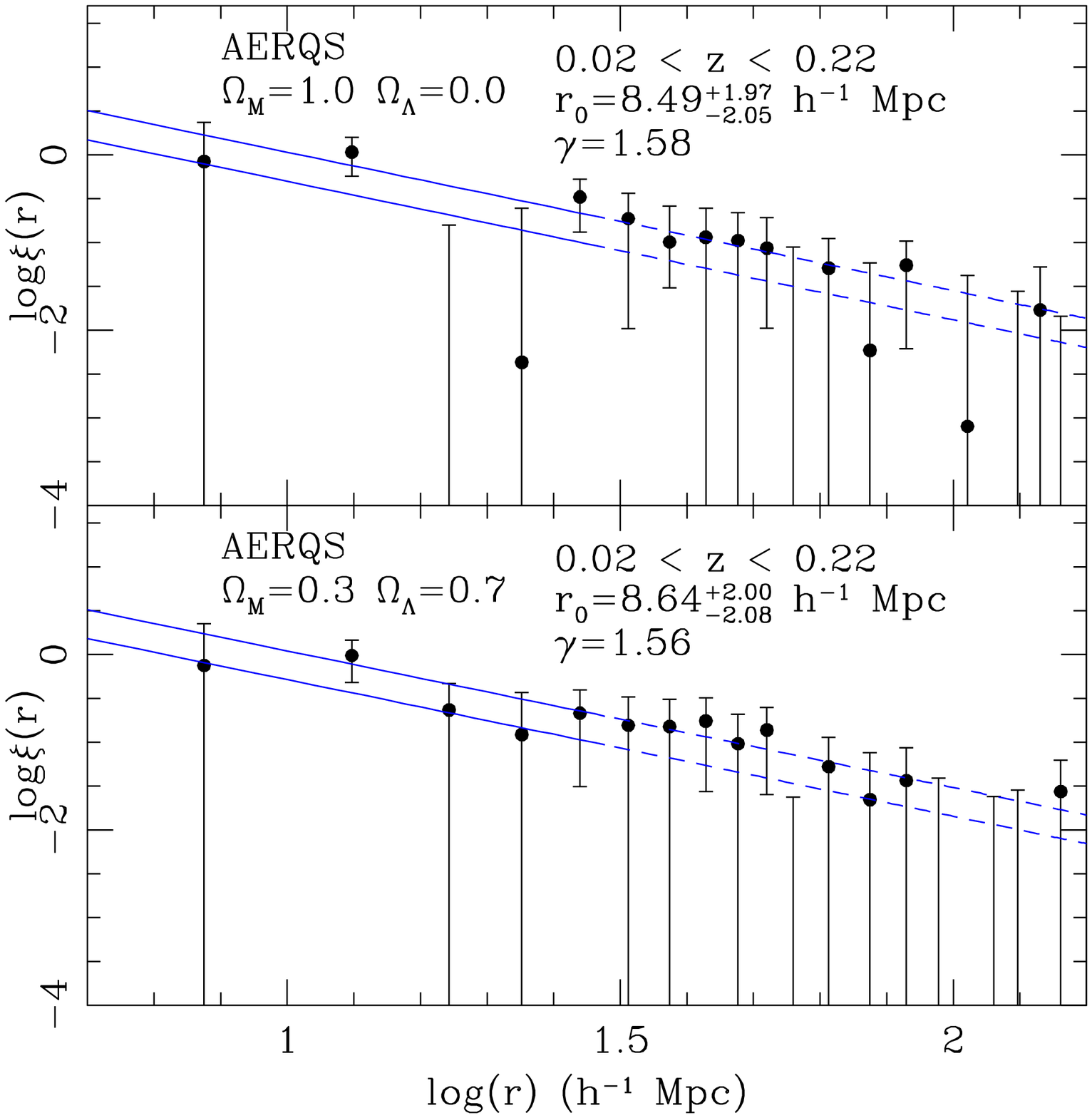}   
\caption{The (differential) two-point correlation function    
(with 1$\sigma$ error bars) for the AERQS sample in EdS (top panel)  
and $\Lambda$ models (bottom panel).  The solid lines show the  
1$\sigma$ confidence region obtained by fitting the data with a  
power-law relation (with fixed slope) using the MLE approach described  
in the text. Only points at $r\le 30 h^{-1}$ Mpc are used to fit the TPCF, 
and they are indicated by the solid line. The dashed line indicates the 
extension of the TPCF relation to data that are not used for the fit. 
At small scales ($\log(r) < 1.8$) a bin size of $5 h^{-1}$ Mpc 
has been adopted, while at larger scales ($\log(r)\ge 1.8$) a bin 
sizes of $10 h^{-1}$ Mpc has been used. The bin size was chosen too avoid 
too large bin to bin fluctuations.} 
\label{totlog}
\end{figure}

The differential TPCF for the complete AERQS sample is shown in 
Fig. \ref{totlog}, both for EdS (upper panel) and $\Lambda$ models 
(lower panel). The results have been fitted by adopting a power-law   
relation 
\begin{equation}   
\xi(r)=\left(\frac{r}{r_{0}}\right)^{-\gamma}\ .   
\label{eq:powlaw}   
\end{equation}   
   
The best-fit parameters can be obtained by using a maximum likelihood   
estimator (MLE) based on Poisson statistics and unbinned data   
\citep{croft}.  Unlike the usual $\chi^2$-minimization, this method   
allows to avoid the uncertainties due to the bin size (see above), the 
position of the bin centers and the bin scale (linear or logarithmic). 
   
To build the estimator, it is necessary to estimate the predicted
probability distribution of quasar pairs, given a choice for the
correlation length $r_0$ and the slope $\gamma$. The small number of
pairs observed at small scales makes a reliable determination of the
slope $\gamma$ particularly difficult. Therefore, we have used fixed
values for the slope $\gamma$, adopting those obtained by
\citet{2qzclust} for the 2dF catalog, namely $\gamma=1.58$ and
$\gamma=1.56$ for EdS and $\Lambda$ models, respectively.  In this
way, the comparison with the TPCF at higher redshifts obtained from
the 2dF data is equivalent both in term of $r_0$ and $\bar\xi$.
 
By using all the distances between the   
quasar-random pairs, we can compute the number of pairs $g(r)dr$ in   
arbitrarily small bins $dr$ and use it to predict the mean number of   
quasar-quasar pairs $h(r)dr$ in that interval as   
 
\begin{equation}   
h(r)dr=   
{{N_c-1}\over{2N_r}} [1+\xi(r)] g(r)dr\ ,   
\end{equation}   
where the correlation function $\xi$ is modeled with a power-law as in   
Eq.(\ref{eq:powlaw})\footnote{Actually the previous equation holds only for   
the \citet{dp83} estimator (the original formulation for the   
TPCF, $\xi(r)=\frac{QQ(r)}{RR(r)}-1$), but,   
since the results obtained using different estimators are similar, we   
can safely apply it here}. In this way, it is possible to use all 
the distances between the $N_p$ quasar-quasar pairs data to build 
a likelihood. In 
particular, the likelihood function ${\cal L}$ is defined as the   
product of the probabilities of having exactly one pair at each of the   
intervals $dr$ occupied by the quasar-quasar pairs data and the   
probability of having no pairs in all other intervals. Assuming a   
Poisson distribution, one finds   
\begin{equation}   
{\cal L}= \prod_i^{N_p} \exp[-h(r)dr] h(r)dr \prod_{j\ne i}   
\exp[-h(r)dr]\ ,   
\end{equation}   
where $j$ runs over all the intervals $dr$ where there are no pairs.   
It is convenient to define the usual quantity $S=-2 \ln {\cal L}$,   
which can be written, once we retain only the terms depending on the   
model parameter $r_0$, as   
\begin{equation}   
S=2\int^{r_{\rm max}}_{r_{\rm min}} h(r)dr -2\sum_i^{N_p}    
\ln[h(r_i)]\ .   
\end{equation}   
  
The integral in the previous equation is computed over the range of
scales where the fit is made. The minimum scale is set by the smallest
scale at which we find QSO pairs ($r_{\rm min}=3 h^{-1}$ Mpc), while
for the maximum scale we adopt $r_{\rm max}=30 h^{-1}$ Mpc.  The
latter choice is made to avoid possible biases from large angular
scales, where the signal is weak.
 
By minimizing $S$ one can obtain the best-fitting parameter $r_0$. The   
confidence level is defined by computing the increase $\Delta S$ with   
respect to the minimum value of $S$. In particular, assuming that   
$\Delta S$ is distributed as a $\chi^2$  with one degree   
of freedom, $\Delta S=1$ corresponds to 68.3 per cent confidence   
level. It should be noted that by assuming a Poisson distribution 
the method considers all pairs as independent, neglecting their   
clustering. Consequently the resulting error bars can be   
underestimated [see the discussion by \citep{croft}]. 
 
In Fig. \ref{totlog} the lines represent the 1$\sigma$ confidence
region computed with the MLE method previously described, varying only
the correlation length $r_0$.  We find $r_0=8.49 ^{+1.97}_{-2.05}
h^{-1}$ Mpc for the EdS model (with $\gamma=1.58$) and $r_0=8.64
^{+2.00}_{-2.08} h^{-1}$ Mpc for the $\Lambda$ model (with
$\gamma=1.56$).  The quoted errors on $r_0$ are based on the
assumption of a fixed slope. It is well known that the errors on $r_0$
and $\gamma$ are correlated.  Fixing the slopes to $\gamma=1.58$ and
1.56 allows us to derive the confidence levels for the integrated TPCF
$\bar{\xi}$ which can be consistently compared with 2QZ results.
 
It can be useful to present the previous results in a non-parametric  
form, specified by the clustering amplitude within a given comoving  
radius, rather than as a scale length which depends on a power-law fit  
to $\xi(r)$. This is generally represented by the correlation function  
integrated over a sphere of a given radius in redshift-space $r_{\rm  
max}$,  
\begin{equation}   
\bar{\xi}(r_{\rm max})=\frac{3}{r_{\rm max}^3}\int_0^{r_{\rm max}}    
\xi(x)x^2{\rm d}x \ .   
\end{equation}   
This is the same quantity we plotted in Figs. \ref{totflat} and   
\ref{totlambda} for varying $r_{\rm max}$. Different authors have chosen    
a variety of values for $r_{\rm max}$, e.g. $10 h^{-1}$ Mpc   
\citep{sb94,cs96}, $15 h^{-1}$ Mpc \citep{lac98}, or $20 h^{-1}$ Mpc   
\citep{2qzclust}.   
In general, the larger the scale on which the clustering is measured, 
the easier the comparison with the linear theory of the structure 
evolution. Since in the following sections we will compare our results 
with those obtained for the 2QZ by \citet{2qzclust}, we prefer to 
quote clustering amplitudes within $20 h^{-1}$ Mpc, a scale for which 
linearity is expected to better than a few per cent. Choosing a large 
radius also reduces the effects of small scale peculiar velocities and 
redshift measurement errors, which may well be a function of redshift. 
   
Table \ref{tabclust} \ summarizes the values of $r_0$, $\gamma$ and  
$\bar{\xi}(20)$ for the total sample, both for the EdS and $\Lambda$  
models. In the same table, we list the mean redshift ($\bar{z}$) of  
the observed QSO sample. We also report the median value of the  
redshift of the QSO pairs computed within a sphere of $20 h^{-1}$ Mpc,  
$z_{\xi}$.  We find that it is systematically lower than the mean  
redshift of the sample.  
   
In our analysis, we do not take into account the velocity field   
of QSOs, the cone edge effect and the effect of statistical errors on   
QSO redshifts.  Recent papers [see e.g. \citet{2qzclust}] suggest that   
the Poisson errors, due to the limited size of a sample, are more   
important than these effects.   
   
\section{Comparison with other surveys}   
   
It is instructive to compare the present results on the clustering of
low-$z$ AGN with that of other surveys, both at low- and
high-redshift, in order to get information about the connection
between various galactic structures and their evolution.  To avoid
problems with different assumptions on the values of the slope
$\gamma$, we decided to compare the values of the integrated TPCF at
20 $h^{-1}$ Mpc, $\bar{\xi}(20)$. When not directly available in the
original paper, $\bar{\xi}(20)$ has been computed by integrating the
TPCF with the best fitting values of $r_0$ and $\gamma$.
 
\subsection{Comparison with other local AGN surveys}   
   
Using a low-redshift ($z\le 0.2$) sample, \citet{bm93} measured the
clustering properties of 183 AGN in the EMSS. They found evidence for
a small value of the integrated TPCF, $\bar{\xi}=0.7\pm 0.6$ (computed
at $10h^{-1}$ Mpc), corresponding to a correlation length of
$r_0=5.0^{+1.9}_{-3.3} h^{-1}$ Mpc.  The assumed slope for the TPCF is
$\gamma=1.8$ and the resulting $\bar{\xi}(20)$ is $0.20\pm
0.17$. Considering the uncertainties, this result is slightly lower
than or consistent with our results. Moreover, since the Boyle \& Mo
sample is obtained by identifications of X-ray sources, it contains
fainter\footnote{AGN in the EMSS are typically 5 times fainter than
$L^*$ at $z\sim 0.2$, or 1.75 magnitude fainter than $M^*_B$.}  AGN
than AERQS. As a consequence, a slightly smaller value of $r_0$ is
expected for their sample, because the clustering strength is found to
depend, weakly, on the absolute magnitude $M_B$, as shown in
\citet{ximb} and in \citet{2dfgal}.
   
\citet{gs94} investigated the clustering properties of 192 Seyfert galaxies   
from the IRAS All Sky Survey. They claimed a $2-3\sigma$ detection at   
$10-20 h^{-1}$ Mpc, corresponding to $\bar{\xi}(20)=0.14\pm0.15$ 
at $\bar{z}=0.05$, similar to local late-type galaxies. This   
result is consistent with a model in which local QSOs randomly sample   
the galaxy distribution.   
   
\citet{carrera} analyzed the clustering of 235 X-ray selected AGN with   
$0\le z\le 3$, obtaining an integrated TPCF of $0.02\le \bar{\xi}(20) 
\le 0.25$. The redshift range of this survey is particularly extended and the 
density of sources correspondingly low.  The clustering detection is 
marginal, at $2\sigma$ level only. Moreover there are only 33 AGN with 
$z\le 0.2$ in this sample. 
   
\citet{akylas} investigated the angular correlation function of 2096   
sources selected from the RASS-BSC. They rejected known stars and   
other contaminants: a cross-correlation analysis with spectroscopic   
samples indicated that the majority of their sources are indeed   
AGN. They obtained a $\sim 4\sigma$ detection of clustering.  Using   
the Limber equation and assuming a source redshift distribution (not   
shown in their paper) with an estimated mean value of 0.1, they derived   
$\bar{\xi}(20)=0.35\pm 0.09$.  Stars, galaxy clusters or other   
spurious contaminants could affect their results.   
 
\citet{mullis01} derived the clustering properties of 217 AGN found in the   
North Ecliptic Pole (NEP) survey, a connected area of $\sim 81$ 
deg$^2$ covered by ROSAT observations. The sample spans the redshift 
interval $0\le z\le 3.889$, with $\bar{z}=0.408$. A $3.8\sigma$ 
clustering detection was obtained, corresponding to an integrated TPCF 
of $\bar{\xi}(20)=0.36\pm0.15$. This result confirms that X-ray 
selected AGN are spatially clustered in a manner similar to that of 
optically/UV selected AGN. 
   
Notice that \citet{bm93}, \citet{gs94}, \citet{carrera} and   
\citet{akylas} used an EdS cosmology to compute the clustering   
properties of their samples, while \citet{mullis01} adopted a   
$\Lambda$ model.   
 
Finally, it is interesting to compare the clustering properties of AGN 
and normal galaxies at low-$z$, using our results and recent results 
by \citet{2dfgal}. Our value for the AGN correlation strength 
($\bar{\xi}(20)\sim 0.461\pm 0.237$) appears slightly lower than the 
typical value for the brighter early-type galaxies 
($\bar{\xi}(20)=0.70^{+0.11}_{-0.08}$), or at most consistent, 
indicating that these two classes have not experienced a completely 
different evolutionary history, but could represent two distinct 
phases during the processes of formation and evolution of the same 
objects.  This gives an additional support for models dealing with 
the joint evolution of QSOs and normal galaxies [e.g. see 
\citet{hk00,granato,xfranc} and references therein]. In particular, 
the fact that the correlation length of AGN at $z\sim 0$ is consistent 
with that of ellipticals or S0 in the local universe, reinforces 
the hypothesis that the QSO host galaxy should be old. 
 
\subsection{Comparison with QSO clustering at high-redshift}   
   
\citet{lac98} investigated the evolution of QSO clustering using a sample of   
388 QSOs with $0.3\le z\le 2.2$ over a connected area of 25 deg$^2$ 
down to $B\le 20.5$ magnitude. Evidence was found for an increase of 
the clustering with increasing redshift 
($\bar{\xi}(20)=0.22^{+0.28}_{-0.18}$ at $0.3\le z\le 1.4$ and 
$\bar{\xi}(20)=0.87^{+0.38}_{-0.21}$ at $1.4\le z\le 2.2$). This 
result does not support the idea of a single population model for 
QSOs. The general properties of the QSO population studied by 
\citet{lac98} would arise naturally if QSOs are short-lived events 
($\tau \sim 10^{6}-10^{7}$ yr) related to a characteristic halo mass 
of $\sim 5\cdot 10^{12} M_{\odot}$. 
   
\begin{figure}   
\epsscale{.8}   
\plotone{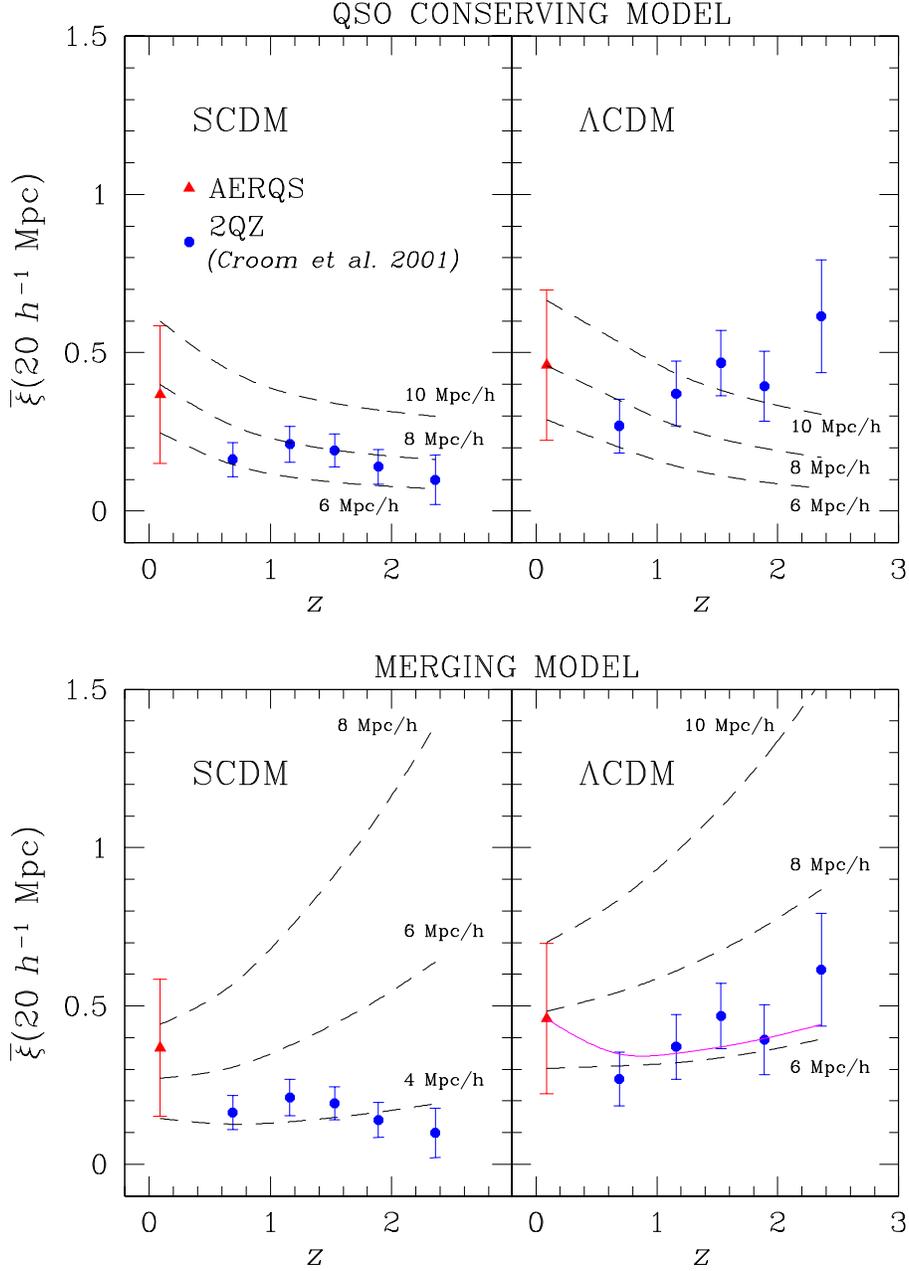}   
\caption{The evolution of the integrated TPCF, $\bar{\xi}(20)$.   
This plot compares the observational data, shown by points with  
1$\sigma$ error bars, to the predicted redshift evolution of the  
clustering for QSO-conserving (top panels) and merging (bottom  
panels). The point at $z=0.1$ is our AERQS result, while points at  
higher redshifts come from 2QZ analysis.  Theoretical results for SCDM  
and $\Lambda CDM$ are shown in the left and right panels,  
respectively. The dashed lines represent the clustering evolution for  
models with a given value of the correlation length $r_0$ at $z=0$, as  
indicated in the plot. The solid line in the bottom-right panel refers  
to the predictions of a combined model, which assumes a merging phase  
(with $\log M_{\rm min}=12.5$) at high-$z$ and a following conserving  
phase at low-$z$ (see the text for more details).  
\label{conserving}}   
\end{figure}

\citet{2qzclust} have used more than 10,000 QSOs taken from the    
preliminary data release catalog of 2QZ to measure the QSO clustering 
as a function of redshift. Their sample spans two connected areas for 
a total of 750 deg$^2$ at a limiting magnitude of 20.85 in the $B_J$ 
band.  The completely identified sample (not yet released) consists of 
nearly 22,500 QSOs in the redshift range $0.3\le z\le 2.2$.  The 
results from the preliminary data release (to be considered with some 
caution), expressed in terms of the correlation function integrated 
inside spheres of $20h^{-1} $ Mpc, $\bar{\xi}(20)$, are shown in 
Fig. \ref{conserving}, together with the estimates obtained 
at $z\sim 0.1$ for the AERQS sample.  The discussion of the theoretical 
models shown by the different lines, will be given in the following 
Section. 
  
For an EdS universe (left panels), \citet{2qzclust} find that there is 
no significant evolution of the QSO clustering in comoving coordinates 
over the whole redshift range considered.  Assuming a $\Lambda$ model 
(right panels), the clustering shows a marginal increase at high 
redshifts, with a minimum of $\xi(r)$ near $z\sim 0.5-1.0$.  Our data 
show a tendency to an increase of $\bar{\xi}(20)$ at low-$z$, both for 
the EdS and $\Lambda$ models.  This result supports the predictions 
based on simple theoretical models (see next section) and on numerical 
simulations by \citet{bagla98a}.  Notice that very recently this 
general trend for the clustering evolution has been also confirmed by 
the power spectrum analysis made by \citet{outram} using the final 
version of the 2QZ catalog, containing 22,652 QSOs. 
 
The AERQS AGN catalog samples a part of the QSO luminosity function
which is fainter than that sampled by the 2QZ. The mean absolute
magnitudes of the total AERQS QSOs are $M_B=-23.49$ and -22.99 for the
EdS and $\Lambda$CDM, respectively. The 2QZ QSOs at $0.3\le z\le 2.2$
are brighter than local AGN, with $M_J=-24.43$ and -25.11 for EdS and
$\Lambda$, respectively.  In our comparison, we do not take into
account the dependence of the TPCF on the absolute magnitude of the
sample, since the QSO population exhibits a strong luminosity
evolution with redshift and \citet{ximb} have demonstrated that the
dependence of the clustering on $M_B$ is very weak.  A correction of
the luminosity dependence of the TPCF would increase the AERQS value,
enhancing the redshift evolution of the clustering.
  
\section{Modeling the redshift evolution of QSO clustering}   
   
By adding the AERQS value of local QSO correlation length to the 2QZ   
estimates at higher redshifts, we have now the complete picture of the   
QSO clustering properties up to $z\sim 2.5$, as summarized in   
Fig. \ref{conserving}. In the following, we introduce a model 
which can be used to interpret the observed evolution.   
   
In general, the theoretical understanding of how matter clustering   
grows via gravitational instability in an expanding universe is presently 
quite well developed, even if the number of ingredients required in   
the models is large.  As a consequence, it is relatively   
straightforward to compute the correlation function of matter   
fluctuations, $\xi_{\rm m}$, as a function of redshift, given a   
cosmological scenario [see e.g. \citet{pd96,smith03}].  However, this   
does not lead directly to a prediction of QSO correlation properties   
because the details of the link between the distribution of active   
nuclei and the distribution of the mass are not fully understood. In   
principle, this relationship could be highly complex, non-linear and   
environment-dependent, making very difficult to 
obtain useful informations on the evolution of matter fluctuations 
from the AGN clustering. In this spirit, a   
relatively simple form of the local bias $b$ is generally assumed. 
   
Matarrese et al. (1997; see also Moscardini et al. 1998; 
Hamana et   
al. 2001) developed an algorithm for describing the clustering on our   
past light-cone taking into account both the non-linear dynamics of   
the dark matter distribution and the redshift evolution of the bias   
factor. The final expression for the observed spatial correlation   
function $\xi_{\rm obs}$ in a given redshift interval ${\cal Z}$ is   
\begin{equation}   
\xi_{\rm obs}(r) = { \int_{\cal Z}   
d z_1 d z_2 {\overline{\cal N}}(z_1) {\overline{\cal N}}(z_2)   
~b_{\rm eff}(z_1) b_{\rm eff}(z_2)   
\xi_{\rm m}(r,\bar{z})   
\over \bigl[ \int_{\cal Z} d z_1   
{\overline{\cal   
N}}(z_1) \bigr]^2 } \;,   
\label{eq:xifund}   
\end{equation}   
where ${\overline{\cal N}}(z)\equiv {\cal N}(z)/r(z)$, ${\cal N}(z)$   
is the actual redshift distribution of the catalog and $r(z)$   
describes the relation between comoving radial coordinate and   
redshift.  Here $\bar{z}$ is a suitably defined intermediate redshift.   
The method has been extended to include the effects of redshift-space  
distortions using linear theory and the distant-observer approximation  
\citep{kaiser}.  
  
A fundamental role in the previous equation is played by the effective  
bias $b_{\rm eff}$. In fact the final aim of models dealing with  
clustering is to determine the behavior of the bias factor, once a  
given theoretical picture is assumed. In practice the effective bias  
can be expressed as a weighted average of the `monochromatic' bias  
factor $b(M,z)$ of objects with some given intrinsic property $M$  
(like mass, luminosity, etc):  
\begin{equation}   
b_{\rm eff}(z) \equiv {\cal N}(z)^{-1} \int_{\cal M} d\ln M' ~b(M',z)   
~{\cal N}(z,M')\, ,   
\label{eq:b_eff}   
\end{equation}   
where ${\cal N}(z,M)$ is the number of objects actually present in the   
catalog with redshift within $dz$ of $z$ and property within $d\ln M$ of   
$\ln M$, whose integral over $\ln M$ is ${\cal N}(z)$.   
   
In most fashionable models of structure formation, the growth of   
large-scale features happens because of the hierarchical merging of   
sub-units.  Since the development of the clustering hierarchy is   
driven by gravity, the most important aspects to be understood are the   
properties of dark halos rather than the QSOs residing in them.   
Following \citet{mw96}, it is possible to calculate the bias   
parameter $b(M,z)$ for halos of mass $M$ and `formation redshift'   
$z_f$ observed at redshift $z\leq z_f$ in a given cosmological model as   
\begin{equation}   
b(M,z\vert z_f) = 1 + {1\over \delta_c} {D_+(z_f) \over D_+(z)}   
\biggl( {\delta_c^2 \over \sigma_M^2 D_+(z_f)^2 } - 1\biggr) \;,   
\label{eq:b_mono}   
\end{equation}   
where $\sigma^2_M$ is the linear variance averaged over the scale  
corresponding to the mass $M$, extrapolated to the present time  
($z=0$); $\delta_c$ is the critical linear over-density for spherical  
collapse; $D_+$ is the growing factor, depending on the cosmological  
parameters $\Omega_M$ and $\Omega_\Lambda$. The distribution in  
redshift and mass $\bar n(z, M)$ for the dark halos can be estimated  
using the \citet{ps74} formalism; in particular in the following  
analysis we adopt the relation found by \citet{st99}. In the standard  
treatment of hierarchical clustering, {\em all} the halos that exist  
at a given stage merge immediately to form higher mass halos, so that  
in practice at each time the only existing halos at all are those  
which just formed at that time (i.e. $z_f=z$). If one identifies  
quasars with their hosting halos, then the merging rate is  
automatically assumed to be much faster than the cosmological  
expansion rate. This is at the basis of what \citet{mclm97} and  
\citet{mclm98} called {\em merging model}.  Of course this  
instantaneous-merging assumption is physically unrealistic and is  
related to the fact that we use a continuous mass variable, while the  
aggregates of matter that form are discrete.  Assuming a monotone  
relation between the mass and the observational quantity defining the  
limits of a given survey, the effective bias can be estimated by  
considering that the observed objects represent all halos exceeding a  
certain cutoff mass $M_{\rm min}$ at any particular redshift. In this  
way, by modeling the linear bias at redshift $z$ for halos of mass $M$  
as in equation (\ref{eq:b_mono}) and by weighting it with the  
theoretical mass--function $\bar n(z, M)$, which can be  
self--consistently calculated using the \citet{st99} relation, the  
behavior of $b_{\rm eff}(z)$ is obtained.  The parameter $M_{\rm min}$  
can be regarded as a free parameter or alternatively fixed in order to  
obtain given values of the correlation length $r_0$ at $z=0$ (see  
later).  
    
An alternative picture of biasing can be built by imagining that   
quasar formation occurs at a relatively well-defined redshift   
$z_f$. Actually there are no changes if one assume that there is some   
spread in the distribution of $z_f$.  If this is the case, one can   
further imagine that quasars, which are born at a given epoch $z_f$,   
might well be imprinted with a particular value of $b(M,z_f)$ as long   
as the formation event is relatively local. If quasars are biased by   
birth in this way, then they will not continue with the same biasing   
factor for all time, but will tend to be dragged around by the   
surrounding density fluctuations, which are perhaps populated by   
objects with a different bias parameter. In this case, the evolution   
of the bias factor can be obtained from \citep{fry96}:   
\begin{equation}   
b(z)= 1+ (b_f-1) { D_+(z_f)  \over D_+(z) } \;, \\\\\\\\\\ ~z<z_f \;,   
\label{eq:b_cons}   
\end{equation}   
where $b_f$ is the bias at the formation redshift $z_f$.  Notice that 
$b(z)$ approaches unity with time, provided that the universe does not 
become dominated by curvature or vacuum in the meantime \citep{cmp98}. 
This model is called {\em conserving model} [see \citet{mclm97} and 
\citet{mclm98}] or, alternatively, {\em test particle model}.  Again, 
it is difficult to motivate this model in detail because it is hard to 
believe that all galaxies survive intact from their birth to the 
present epoch, but at least it gives a plausible indication of the 
direction in which one expects $b$ to evolve if the timescale for 
quasar formation is relatively short and the timescale under which 
merging or disruption occurs is relatively long. 
   
Notice that the merging model (rapid merging) and conserving model (no   
merging) can be regarded as two extreme pictures of how structure   
formation might proceed.  In between these two extremes, one can   
imagine more general scenarios in which quasars neither survive   
forever nor merge instantaneously. The price for this greater   
generality is that one would require additional parameters to be   
introduced in the models (see the discussion at the end of the next   
section).   
   
\subsection{Results}   
   
In the following analysis we will present the results for two  
different cosmological models.  Both models assume a CDM power  
spectrum \citep{bardeen}, with spectral index $n=1$ and shape parameter  
$\Gamma=0.2$.  The power spectrum normalization (expressed in terms of  
$\sigma_8$, i.e. the r.m.s. fluctuation amplitude in a sphere of  
$8h^{-1}$ Mpc) is chosen to be consistent with very recent estimates  
obtained from the cluster abundance analysis [e.g. \citet{rb02, vnl02,  
seljak02}].  
The two considered models  
are:  
\begin{itemize}   
\item a ``standard'' CDM Einstein-de Sitter model with    
$\sigma_8=0.5$ (hereafter SCDM);    
\item a  flat CDM universe with    
($\Omega_{M}$,$\Omega_{\Lambda})=(0.3,0.7)$ with $\sigma_8=0.8$  
(hereafter $\Lambda CDM$).  
\end{itemize}   
   
\begin{figure}   
\epsscale{1.05}   
\plotone{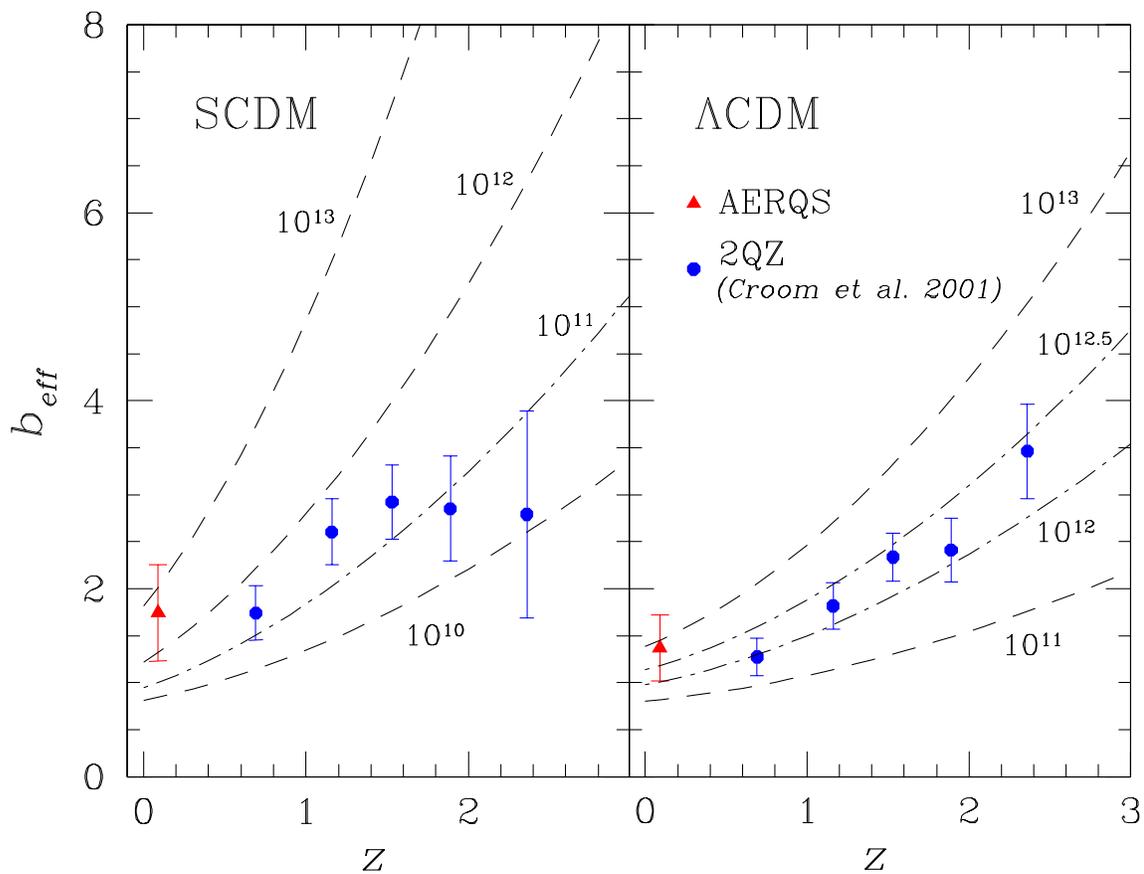}   
\caption{The redshift evolution of the QSO bias factor. The points with   
1$\sigma$ error bars represent the observational estimates: the point  
at $z=0.1$ comes from this analysis of the AERQS catalog, while the  
remaining data are from 2QZ.  The lines show the evolution of the bias  
obtained assuming different values of the mass $M_{\rm min}$ (in units  
of $h^{-1} \msun$).  
\label{transient}}   
\end{figure}

In Fig. \ref{transient} we show the redshift evolution of the bias for  
different values (indicated in the plot in units of $ h^{-1}  
\msun$) of the minimum mass $M_{\rm min}$ of the halos hosting QSOs.    
The theoretical predictions are compared to the observational results,  
shown by the points with 1$\sigma$ error bars. The values at $z=0.1$  
represent the bias parameter derived by our analysis of the AERQS.  
They are obtained by dividing the measured integrated TPCF for QSOs,  
${\bar\xi}(20)$ by the theoretically predicted autocorrelation  
function of the underlying matter $\bar{\xi}_{\rm m}(20)$:  
$b^2={\bar\xi}(20)/\bar{\xi}_{\rm m}(20)$.  We obtain $b=1.75\pm 0.51$  
and $b=1.37\pm 0.35$ for SCDM and $\Lambda CDM$, respectively.  The  
data for higher redshifts come from the analysis of the 2QZ survey  
\citep{2qzclust}.  
   
A first comparison shows that the values for AERQS are consistent with   
the values at $z\sim 0.7$ for the 2QZ, implying the absence of a   
significant evolution of bias at low redshifts. As already noticed by   
\citet{2qzclust}, the trend at higher redshifts for bias appears in   
general to depend on the cosmological models: for $\Lambda CDM$ model,   
the observed $b$ is always an increasing function of redshift, while   
in the SCDM case the value of $b$ is almost constant for $z\ge 1.5$.   
   
More interesting is the comparison of the observed $b$ with the 
theoretical predictions obtained assuming different $M_{\rm min}$. For 
the $\Lambda CDM$ model the AERQS value corresponds to $\log 
M_{\rm min}=12.7^{+0.8}_{-0.7}$ (1$\sigma$ error bars), and the 
observed trend is consistent with the bias evolution expected for dark 
halos with a minimum mass almost constant in redshift ($\log M_{\rm 
min}\sim 12-12.5$). On the contrary, for the SCDM model it is 
impossible to reproduce the bias factor using a constant minimum mass: 
while the value for AERQS suggests $\log M_{\rm min}=12.5\pm{0.7}$ 
(always 1$\sigma$ error bars), the bias factor corresponds to halos 
with $\log M_{\rm min}\sim 11.5$ at $0.5 \le z\le 1.5$ and $\log 
M_{\rm min}\le 11.5$ at $ z\ge 2$. 
   
In Fig. \ref{conserving} we show the predictions for the redshift  
evolution of the TPCF integrated over $20 h^{-1}$ Mpc computed  
adopting the QSO-conserving (upper panels) and merging (bottom panels)  
models, described above. The points (with 1$\sigma$ error bars) refer  
to the observational estimates, again from AERQS at $z=0.1$ and from  
2QZ at higher redshifts.  The dashed lines represent the results  
obtained for models built to have given values of the QSO correlation  
length $r_0$ at $z=0$. In particular in the case of the  
QSO-conserving model, we show results for $r_0(z=0)=6, 8, 10$ $h^{-1}$  
Mpc for both models. In the case of the merging model, we show the  
results for $r_0(z=0)= 4, 6, 8 h^{-1}$ Mpc, corresponding to a minimum  
dark matter halo mass of $3.4\times 10^{11}, 4.6\times 10^{12}, 1.7  
\times 10^{13} h^{-1} \msun$, for SCDM, and for $r_0(z=0)= 6, 8, 10   
h^{-1}$ Mpc, corresponding to $M_{\rm min}=2.5\times 10^{12},1.2  
\times 10^{13} h^{-1}, 2.9 \times 10^{13} h^{-1} \msun$ for $\Lambda  
CDM$.  
   
From the figure, it is evident that in the case of SCDM the 
QSO-conserving model is more or less able to reproduce the clustering 
evolution over the whole redshift interval, once a local value of 
$r_0\sim 7 h^{-1}$ Mpc is used. The situation is quite different for 
$\Lambda CDM$, for which the high clustering observed at $z\ge 2$ is 
not compatible with any trend predicted by the QSO-conserving model: 
only for $z\le 1$ the decrease of $\bar{\xi}(20)$ follows the model 
expectations corresponding to a local value of $r_0 (z=0)\sim 7 
h^{-1}$ Mpc. 
   
The comparison of the QSO clustering with the predictions of the   
merging model shows only marginal agreement on the whole interval   
$0.0\le z\le 2.5$.  In particular, for SCDM the observational data are   
close to the predictions of the model corresponding to a low value of   
the local clustering, $r_0 (z=0)\sim 4 h^{-1}$ Mpc, while for $\Lambda   
CDM$ the better agreement is for models corresponding to $r_0   
(z=0)\sim 7 h^{-1}$ Mpc, but with large deviations.   
   
As already said, these simple schemes do not exhaust all the possible 
scenarios through which QSOs might have formed and evolved. For 
example, it is quite possible that merging could play a different role 
at different redshifts. Present-day AGN, for example, have clearly not 
just formed at the present epoch since their observational properties 
suggest a lack of mergers in the recent past. On the other hand, it is 
plausible that QSOs at much higher redshifts, say $z\ge 2$, are 
undergoing merging on the same timescale as the parent halos. This 
suggests the possible applicability of a model where rapid merging 
works at high redshifts, but it ceases to dominate at lower redshifts 
and the bias then evolves by equation (\ref{eq:b_cons}) until now. In 
this context it is interesting to note that, while $b_f$ is a free 
parameter in equation (\ref{eq:b_cons}), it is actually predicted, 
once the appropriate minimum mass is specified. In fact it corresponds 
to the bias at the redshift $z_f$ when objects stop merging.  This 
model has been introduced by \citet{mclm98}, where the resulting 
relations for the redshift evolution of the bias factor are given. The 
complete application of this combined model to the present data on 
quasar clustering is quite difficult because of the size of error 
bars.  Only as an example, for $\Lambda CDM$ we compute the predicted 
$\bar{\xi}(20)$ by assuming a merging model with $\log M_{\rm 
min}=12.5$, followed by a conserving phase. The result, shown as solid 
line in the bottom-right panel of Fig. \ref{conserving}, is in rough 
agreement with the observational data, indicating that the redshift 
$z_f$, where the transition between the two different regimes occurs, 
is located at approximately $z_f=0.8$. Of course, a validation of this 
model requires more robust estimates of the QSO clustering properties. 
   
\section{Discussion}   
  
\subsection{Clustering in the local universe}   
  
As discussed in the previous paragraphs, there are empirical and 
theoretical evidences that QSOs are biased with respect to the matter 
distribution.  The results of \citet{bm93}, \citet{gs94}, 
\citet{carrera}, \citet{akylas} and \citet{mullis01} are consistent,   
within the uncertainties, with the present results.  In general we 
find that, at low-$z$, AGN have a correlation length of $\sim 8 
h^{-1}$ Mpc.  This value is quite similar to the correlation length of 
ellipticals, EROs or RGs at $z\sim 0$ and higher than that of spirals 
or late-type galaxies. Assuming the hierarchical clustering paradigm 
and the SCDM model, a local correlation length of $\sim 8.5\pm 2 
h^{-1}$ Mpc corresponds to a population of DMHs with mass larger than 
$\sim 10^{12.5}h^{-1} M_{\odot}$ ($10^{11.8}-10^{13.2}h^{-1} M_{\odot}$ at a 
$1\sigma$ confidence level) and a bias parameter of $b\sim 1.7$. 
The space density of DMHs more massive than this limit is $\rho_{\rm 
DMH}\sim 3.71\cdot 10^{-3} h^{3}$ Mpc$^{-3}$ 
($0.63-19.64\cdot 10^{-3} h^{3}$ Mpc$^{-3}$ at $1\sigma$), 
as obtained by applying 
the Press-Schechter formalism (e.g. Sheth \& Tormen 1999).  The space 
density of bright AGN in the local universe is $\rho_{\rm AGN}\sim 
5.7\cdot 10^{-7} h^{3}$ Mpc$^{-3}$, as inferred by \citet{GC2000} 
using a sub-sample of the AERQS with limiting magnitude $M_B=-22.5$. 
We can thus obtain a rough estimate of the duty-cycle of local AGN, 
$\tau_{\rm AGN}$ using the simple relation 
\begin{equation}   
\tau_{\rm AGN}=\frac{\rho_{\rm AGN}}{\rho_{\rm DMH}}\cdot \tau_{\rm H}\ ,   
\end{equation}   
where $\tau_{\rm H}$ is the Hubble time\footnote{$\tau_{\rm H}$ is 
computed at $z=0.1$ and corresponds to $11.3\cdot 10^9$ yr and 
$13.1\cdot 10^9$ yr for the EdS and $\Lambda$ cosmological models, 
respectively}.  The duty-cycle of AGN at $z\sim 0.1$ turns out to be 
$\tau_{\rm AGN}\sim 1.7\cdot 10^{6}$ yr (the 1$\sigma$ confidence 
region is $3.3\cdot 10^5-1.0\cdot 10^7$ yr). This result is in good 
agreement with the one 
\citet{KH02} obtained by comparing their model to the 2QZ data, and only 
marginally consistent with the value of $\tau_{\rm QSO}\sim 10^7$ yr 
obtained by \citet{shs02} for QSOs at $z\sim 3$. It is worth noting 
that both these papers adopt a $\Lambda$ model and $h=0.7$. 
 
If we assume the relation found by   
\citet{ferrarese},  
namely 
\begin{equation}   
M_{\rm BH}\sim 10^7 M_{\odot}\left(\frac{M_{\rm DMH}}{10^{12}M_{\odot}}   
\right)^{1.65},  
\end{equation} 
it is possible to infer the mass of active BHs $M_{\rm BH}$ at $z\sim 
0.1$.  With the values of the AGN DMH mass estimated in our sample, 
$\log h M_{\rm min}/M_{\odot}=12.5\pm{-0.7}$, we can obtain a rough 
estimate for $M_{\rm BH}$: $6.7\cdot 10^{7}h^{-1}~ M_{\odot}$ 
($4.7\cdot 10^{6}- 9.5\cdot 10^{8}h^{-1} M_{\odot}$ at 1$\sigma$). 
Assuming an absolute magnitude of $M_B=-22.5$ that 
corresponds\footnote{assuming a ratio $L_{\rm Bol}/L_B=10$.}  to a 
bolometric luminosity $L_{\rm Bol}=10^{12} L_{\odot}$, one can infer a 
typical ratio $L/M$ for a local AGN of $1.5\cdot 10^4 
L_{\odot}/M_{\odot}$ ($1.1\cdot 10^3-2.1\cdot 10^5 
L_{\odot}/M_{\odot}$). For comparison, the Eddington value is 
$L/M=3.5\cdot 10^4 L_{\odot}/M_{\odot}$. In this case an efficiency of 
$\eta\equiv L/L_{\rm Edd}\sim 0.4$ ($0.03-6.1$) is derived. 
 
For the $\Lambda CDM$ model, the correlation length is $r_0=8.6\pm 2.0 
h^{-1}$ Mpc, which correspond to a DMH mass of the order $\log M_{\rm 
min}=12.7^{+0.8}_{-0.7}$.  Following the same approach carried out for 
the SCDM model, with an AGN density of $\rho_{\rm AGN}\sim 4.9\cdot 
10^{-7} h^{3}$ Mpc$^{-3}$ and $\rho_{\rm DMH}\sim 8.12\cdot 10^{-4} 
h^{3}$ Mpc$^{-3}$ ($1.28\cdot 10^{-4}-4.31\cdot 10^{-3} h^{3}$ 
Mpc$^{-3}$ at 1$\sigma$) we derive a slightly longer duty cycle of 
$\tau_{\rm AGN}\sim 7.9\cdot 10^{6}$ yr ($1.5\cdot 10^{6}-5.0\cdot 
10^{7}$ yr) and an efficiency of $4.8\cdot 10^3 L_{\odot}/M_{\odot}$ 
($3.3\cdot 10^2-1.0\cdot 10^5 L_{\odot}/M_{\odot}$), corresponding to 
$\eta\sim 0.14$ ($0.01-2.8$, at 1$\sigma$). Due to the 
large error bars on the TPCF, the constraints on the efficiency $\eta$ 
are not stringent, but give nevertheless an indication of the mean value 
of the Eddington ratio for local AGN.
   
AGN in the AERQS sample seem thus to accrete in a sub-Eddington 
regime, lower than the nearly- or super-Eddington accretion generally 
assumed for QSOs at high redshifts.  High-$z$ QSOs are thought to have 
relatively small masses, so their extreme luminosities point to a high 
Eddington ratio (1-10 of the standard value).  The direct 
determination of the Eddington ratio for QSOs at high-$z$ is 
complicated by a number of difficulties, as discussed by 
\citet{wu02}, who suggest that the true value at $z\ge 1$ is uncertain   
and dominated by selection effects.  At $z\sim 0.1$ they obtain a 
value $L_{\rm Bol}/L_{\rm Edd}\sim 0.1$, consistent with our result. 
\citet{bechtold}, using Chandra observations of high-redshift QSOs,    
estimated the BH mass and the Eddington ratio at $3.7\le z\le 6.28$   
and compared it with the value for local AGN. At high-$z$ QSOs possess   
masses of the order of $10^{10}h^{-1}~M_{\odot}$ and are growing at a   
mass accretion rate of 0.1 ${\dot{m}}_{\rm Edd}$. At low-$z$ their   
results are comparable with our values, with $M_{\rm BH}$ between   
$10^8$ and $10^9 h^{-1}~M_{\odot}$ and an Eddington ratio $\eta$ between 
$10^{-2}$ and $10^{-1}$.  From the point of view of the theoretical   
modeling, \citet{ciotti} were able to reproduce the QSO LF   
the mass function of local BHs with an Eddington ratio constant and   
equal to $10^{-1}$ in the redshift range $0\le z\le 4$. \citet{HM00}   
used an Eddington ratio decreasing from $z=4$ to $z=0$ with a typical   
value of $10^{-2}-10^{-3}$ at $z\sim 0$.   
   
\subsection{Interpreting QSO and galaxy clustering}   
   
In a sense, the general picture emerging from the observational data is   
that the clustering evolution for galaxies is similar to the QSO one:   
at low-$z$, the correlation length is decreasing from the local value   
reaching a minimum at $z\sim 1$, then it increases till $z\sim 3-4$.   
As discussed by \citet{arnouts99,arnouts02}, the term ``evolution'' 
has not to 
be considered literally. Given a survey defined by its characteristic   
limiting magnitude and surface brightness, the galaxies observed at   
high-$z$ typically have higher luminosities. Therefore, the intrinsic   
differences of the galaxy properties at different $z$ can mimic an   
evolution, i.e. the evolution measured in a flux-limited survey is not   
only due to the evolution of a unique population but can be due to a   
change of the observed population.  The picture emerging from QSO   
clustering points in the same direction: QSOs are not part of a unique   
population, due to their short duty-cycle, and are intrinsically   
related with galaxy evolution.   
   
At this stage, it is important to discuss how the clustering depends 
on absolute magnitude. Moreover, we should consider how the bias 
factor changes when the catalog selection effects are considered, 
i.e. when the theoretical quantities, as the mass $M$, are substituted 
by the observational ones, such as the luminosity $L$.  Chosen one of 
the previously described models, one will end up with the quantity 
$b(M,z)$ to be understood as `the bias that objects of mass $M$ have 
at redshift $z$'. The effective bias at that redshift can be written 
more precisely as (see also Martini \& Weinberg 2001) 
\begin{equation}   
b_{\rm eff}(z) = N(z)^{-1} \int d \ln L \,\Phi_{\rm obs}(L) b[M(L),z] \;,   
\end{equation}   
where $N(z) = \int d \ln L \,\Phi_{\rm obs}(L)$ and $\Phi_{\rm 
obs}(L)$ is the {\em observed} luminosity function of the catalog, 
i.e. the intrinsic luminosity function multiplied by the catalog 
selection function, which will typically involve a cut in apparent 
magnitude, whatever wave-band is being used. 
   
\citet{ximb} observed a weak trend of the clustering strength with the   
magnitude, brighter objects being more clustered. Such a behavior can 
be understood by focusing on the results obtained in the previous 
paragraph.  The efficiency and the accretion rate for local AGN 
determine the relation between mass and luminosity, and how they 
evolve with redshift. As a simple consequence, for QSOs at different 
epochs, luminosity does not necessarily trace mass. The dependence of 
clustering strength on the luminosity is thus weaker than the one 
expected in theoretical models assuming a fixed $M/L$ ratio. 
   
EROs and RGs at $z\sim 1$ have higher correlation amplitude than  
QSOs at the same redshift; indeed they are following slow or 
passive evolution. Probably at $z\sim 0$ they will plausibly become  
the brightest and most massive galaxies inside the clusters.  QSOs,  
instead, show a different evolution for the TPCF: their behavior is  
consistent with that of a typical merging model at high-$z$ and a  
passive evolution or object-conserving at low-$z$.  
    
\citet{KH02} explored theoretically the possibility of using the   
cross-correlation between QSOs and galaxies, $\xi_{QGal}$, to obtain   
new information on the masses of DMHs hosting QSOs. They used a   
semi-analytical model in which super-massive BHs are formed and fueled   
during major mergers.  The resulting DMH masses can be in principle   
used to estimate the typical QSO life-time.  In current redshift   
surveys, like the 2dFGRS or SDSS, these measurements will constrain   
the life-times of low-$z$ QSOs more accurately than QSO   
auto-correlation function, because galaxies have  much higher a space   
density than QSOs. As a result, $\xi_{QGal}$ can yield information   
about the processes responsible for fueling SMBHs.   
   
\section{Conclusions}   
 
The Asiago-ESO/RASS QSO survey (AERQS), an all-sky complete sample of 
392 spectroscopically identified objects ($B\le 15$) at $z\le 0.3$, 
has been used to carry out an extended statistical analysis of the 
clustering properties of local QSOs. 
 
The AERQS makes it possible to remove present uncertainties about the 
properties of the local QSO population and fix an important zero point 
for the clustering evolution and its theoretical modeling.  With such 
a data-set, the evolutionary pattern of QSOs between the present epoch 
and the highest redshifts is tied down. 
 
On the basis of the (integrated and differential) two-point 
correlation functions, we have detected a $3-4\sigma$ clustering 
signal, corresponding to a correlation length $r_0=8.6\pm 2.0 h^{-1}$ 
Mpc and a bias factor $b=1.37\pm 0.35$ in a $\Lambda CDM$ model. A 
similar value of $r_0$, but corresponding to $b=1.75\pm 0.51$, is 
obtained for an Einstein-de Sitter model, confirming previous analysis 
\citep{bm93,gs94,carrera,akylas,mullis01}.  These results shows that 
low-redshift QSOs are clustered in a similar way to radio galaxies, 
EROs and early-type galaxies, while the comparison with recent results 
from the 2QZ at higher redshifts shows that the correlation function 
of QSOs is constant in redshift or marginally increasing toward low 
redshifts. 
 
This behavior can be interpreted with physically motivated models, 
taking into account the non-linear dynamics of the dark matter 
distribution, the redshift evolution of the bias factor, the past 
light-cone and redshift-space distortion effects. The application of 
these models allows us to derive constraints on the typical mass of 
the dark matter halos hosting QSOs: we have found $\log M_{\rm 
DMH}=12.7^{+0.8}_{-0.7}$ (the mass is units of $h^{-1} M_{\odot}$), 
almost independently of the cosmological model.  Using the abundance 
of dark matter halos with this minimum mass and assuming the relation 
found by \citet{ferrarese} between the masses of dark matter halos and 
active black holes, from the clustering data we can directly infer an 
estimate for the mass of the central active black holes and for their 
life-time, $M_{\rm BH}\sim 2.1\cdot 10^8h^{-1} \msun$ ($1.0\cdot 
10^{7}-2.9\cdot 10^{9}h^{-1} M_{\odot}$) and $\tau_{AGN}\sim 7.9\cdot 
10^6$ yr ($1.5\cdot 10^{6}-5.0\cdot 10^{7}$ yr), respectively.  This 
means that local AGN seem to accrete in a sub-Eddington regime.  All 
these values have been obtained for a $\Lambda CDM$ model; slightly 
shorter duty cycles are derived for an Einstein-de Sitter model.  The 
time-life of $z\sim 3$ QSOs is $\sim 10^7~ yr$, measured by 
\citet{shs02}.  This could be a first indication that QSOs at all 
epochs have a similar life-time, which does not depend strongly on the 
Hubble time. 
 
Observational data suggest that most nearby galaxies contain central 
super-massive black holes, supporting the idea that most galaxies pass 
through a QSO/AGN phase.  However, the different clustering 
properties, together with the the short lifetimes derived for local 
AGN, suggest that this phase picks out a particular time in the 
evolution of galaxies, e.g. epochs of major star formation, 
interactions or merging. In this way the study of the QSO clustering 
evolution using extended catalogs helps us to distinguish between a 
number of possible QSO formation mechanisms. 
 
\acknowledgments  
This work has been partially supported by the European Community
Research and Training Network "Physics of the Intergalactic Medium",
by the Italian MIUR (Grant 2001, prot. 2001028932, ``Clusters and
groups of galaxies: the interplay of dark and baryonic matter''), by
CNR and ASI.  AG was supported by the ESO DGDF 2000 and by an ESO
Studentship and acknowledges the generous hospitality of ESO
headquarters during his stay at Garching.  This project has been also
supported by the European Commission through the ``Access to Research
Infrastructures Action of the Improving Human Potential Programme'',
awarded to the 'Instituto de Astrof\'{\i}sica de Canarias' to fund
European Astronomers access to the European Northern Observatory, in
the Canary Islands.  It is pleasure to warmly thank S. Bianchi,
C. Mullis, P. Andreani, N. Menci and A. Merloni for enlightening
discussions and precious suggestions on the clustering properties of
AGN in the AERQS.  We are grateful to the anonymous referee for useful
comments which improved the presentation of our results.  This paper
makes use of the ROSAT All Sky Survey Bright Source Catalog (1RXS).


\begin{thebibliography}{}   
   
 \bibitem[Akylas et al.(2000)]{akylas} Akylas, A., Georgantopoulos, I. \&   
 Plionis, M. 2000 \mnras~ 318, 1036   
   
 \bibitem[Andreani \& Cristiani(1992)]{ac92} Andreani, P. \& Cristiani, S.   
 1992 \apj~ 398, 13   
   
 \bibitem[Andreani et al.(1994)]{andreani} Andreani, P., Cristiani, 
 S., Lucchin, F., Matarrese, S. \& Moscardini, L. 1994 \apj~ 430, 458 
 
 \bibitem[Arnouts et al.(1999)]{arnouts99} Arnouts, S., Cristiani, S.,   
 Moscardini, L., Matarrese, S., Lucchin, F., Fontana, A. \& Giallongo, E.   
 1999 \mnras~ 310, 540   
   
 \bibitem[Arnouts et al.(2002)]{arnouts02} Arnouts, S., Moscardini, L.,   
 Vanzella, E., Colombi, S., Cristiani, S., Fontana, A., Giallongo, E.,   
 Matarrese, S. \& Saracco, P. 2002 \mnras~ 329, 355   
   
 \bibitem[Bagla(1998)]{bagla98a} Bagla, J. S. 1998 \mnras~ 297, 251   
   
 \bibitem[Bagla(1998b)]{bagla98b} Bagla, J. S. 1998 \mnras~ 299, 417   
   
 \bibitem[Bardeen et al.(1986)]{bardeen}  
 Bardeen, J.M., Bond, J.R., Kaiser, N. \& Szalay A.S. 1986,  
 \apj~  304, 15  
  
 \bibitem[Bechtold et al.(2003)]{bechtold} Bechtold, J.,  
 Siemiginowska, A., Shields, J., Czerny, B., Janiuk, A., Hamann, F.,  
 Aldcroft, T. L., Elvis, M. \& Dobrzycki A.  2003 \apj~ 588, 119   
 
 \bibitem[Boyle \& Mo(1993)]{bm93} Boyle, B. J. \& Mo, H. J.   
 1993 \mnras~ 260, 925   
   
 \bibitem[Carrera et al.(1998)]{carrera} Carrera, F. J., Barcons, X.,   
 Fabian, A. C., Hasinger, G., Mason, K. O., McMahon, R. G., Mittaz, J. P. D.   
 \& Page, M. J. 1998 \mnras~ 299, 229   
 
 \bibitem[Catelan et al.(1998)]{cmp98} Catelan, P., Matarrese, S. \& 
 Porciani, C. 1998 \apj~ 502, L1 
   
 \bibitem[Ciotti et al.(2003)]{ciotti} Ciotti, L., Haiman, Z. \&   
 Ostriker, J. P. in Proceedings of the ESO Workshop  ``The  
 Mass of Galaxies at Low and High Redshift''. R. Bender and A. 
 Renzini, eds,  p. 106 
   
 \bibitem[Croft et al.(1997)]{croft} Croft, R. A. C., Dalton, G. B.,   
 Efstathiou, G., Sutherland, W. J. \& Maddox, S. J., 1997 \mnras~ 291, 305   
   
 \bibitem[Croom \& Shanks(1996)]{cs96} Croom, S. M. \& Shanks, T. 1996   
 \mnras~ 281, 893   
   
 \bibitem[Croom et al.(2001)]{2qzclust} Croom, S. M., Shanks,   
 T., Boyle, B. J.,   
 Smith, R. J., Miller, L., Loaring, N. S. \& Hoyle, F. 2001 \mnras~ 325, 483   
   
 \bibitem[Croom et al.(2002)]{ximb} Croom, S. M., Boyle, B. J.,  
 Loaring, N. S., Miller, L., Outram, P. J., Shanks, T. \& Smith, R. J.  
 2002 \mnras~ 335, 459  
   
 \bibitem[Daddi et al.(2001)]{daddi01} Daddi, E., Broadhurst, T.,   
 Zamorani, G., Cimatti, A., R\"ottgering, H. \& Renzini, A.   
 2001 \aap~ 376, 825   
   
 \bibitem[Daddi et al.(2002)]{daddi02} Daddi, E., Cimatti, A.,  
 Broadhurst, T.,   
 Renzini, A., Zamorani, G., Mignoli, M., Saracco, P., Fontana, A.,   
 Pozzetti, L., Poli, F., Cristiani, S., D'Odorico, S., Giallongo, E.,   
 Gilmozzi, R. \& Menci, N. 2002 \aap~ 384, 1   
   
 \bibitem[Davis \& Peebles(1983)]{dp83} Davis, M. \& Peebles, P. J. E.   
 1983 \apj~ 267, 465   
   
 \bibitem[Ferrarese(2002)]{ferrarese} Ferrarese, L. 2002  
 \apj~ 578, 90   
   
 \bibitem[Franceschini et al.(2002)]{xfranc} Franceschini, A., Braito, V.  
 \& Fadda, D. 2002 \mnras~ 335, 51  
  
 \bibitem[Fry(1996)]{fry96} Fry, J. N. 1996   
 \apj~ 461, L65   
  
 \bibitem[Gehrels(1986)]{gehrels} Gehrels, N. 1986 \apj~ 303, 346  
  
 \bibitem[Georgantopoulos \& Shanks(1994)]{gs94} Georgantopoulos, I. \&   
 Shanks, T. 1994 \mnras~ 271, 773   
   
 \bibitem[Granato et al.(2001)]{granato} Granato, G. L., Silva, L.,  
 Monaco, P., Panuzzo, P., Salucci, P., De Zotti, G. \& Danese, L.  
 2001 \mnras~ 324, 757  
  
 \bibitem[Grazian et al.(2000)]{GC2000} Grazian, A., Cristiani, S.,   
 D'Odorico, V., Omizzolo, A. \& Pizzella, A. 2000 \aj~ 119, 2540; Paper I   
   
 \bibitem[Grazian et al.(2002)]{DSS2k2} Grazian, A., Omizzolo, A.,   
 Corbally, C., Cristiani, S., Haehnelt, M. G. \& Vanzella, E.   
 2002 \aj~ 124, 2955; Paper II    
  
 \bibitem[Haehnelt \& Kauffmann(2000)]{hk00} Haehnelt, M. G. \&  
 Kauffmann, G. 2000 \mnras~ 318, 35  
   
 \bibitem[Haiman \& Menou(2000)]{HM00} Haiman, Z. \& Menou, K.   
 2000 \apj~ 531, 42   
 
 \bibitem[Hamana et al.(2001)]{hyse01}Hamana, T., Yoshida, N., Suto,   
 Y. Evrard, A.E. 2001 \apj~ 561, L143   
    
 \bibitem[Iovino \& Shaver(1988)]{is88} Iovino, A. \& Shaver, P. A. 1988   
 \apj~ 330, 13   
   
 \bibitem[Kaiser(1987)]{kaiser} Kaiser, N. 1987, \mnras~ 227, 1  
  
 \bibitem[Kauffmann \& Haehnelt(2002)]{KH02} Kauffmann, G. \& Haehnelt, M. G.   
 2002 \mnras~ 332, 529   
   
 \bibitem[La Franca \& Cristiani(1997)]{LFC97} La Franca, F. \& Cristiani, S.   
 1997 \aj~ 113, 1517   
   
 \bibitem[La Franca et al.(1998)]{lac98} La Franca, F., Andreani, P. \&   
 Cristiani, S. 1998 \apj~ 497, 529   
   
 \bibitem[Landy \& Szalay(1993)]{ls93} Landy, S. D. \& Szalay, A.S.,  
 1993 \apj~ 412, 64   
 
 \bibitem[Larson(1975)]{larson} Larson, R. B. 1975 \mnras~ 173, 671   
   
 \bibitem[Lynden-Bell(1964)]{lynden} Lynden-Bell, D. 1964 \apj~ 139, 1195   
   
 \bibitem[Martini \& Weinberg(2001)]{mw01} Martini, P.\& Weinberg, D. H. 2001   
 \apj~ 547, 12   
   
 \bibitem[Matarrese et al.(1997)]{mclm97} Matarrese, S., Coles, P.,   
 Lucchin, F. \& Moscardini, L. 1997 \mnras~ 286, 115   
   
 \bibitem[Matteucci et al.(1998)]{matteucci} Matteucci, F., Ponzone, R. \&   
 Gibson, B. K. 1998 \aap~ 335, 855   
   
 \bibitem[Mo \& Fang(1993)]{mf93} Mo, H. J. \& Fang, L. Z. 1993 \apj~ 410, 493    
 \bibitem[Mo \& White(1996)]{mw96}Mo, H. J. \& White, S. D. M.    
 1996 \mnras~ 282, 347   
   
 \bibitem[Moscardini et al.(1998)]{mclm98} Moscardini, L., Coles, P.,   
 Lucchin, F. \& Matarrese, S., 1998 \mnras~ 299, 95   
   
 \bibitem[Mullis et al.(2001)]{mullis01} Mullis, C. R., Henry, J. P.,   
 Gioia, I. M., Boehringer, H., Briel, U. G., Voges, W. \& Huchra, J. P.   
 2001 AAS Meeting 199, 138.19   
   
 \bibitem[Norberg et al.(2002)]{2dfgal} Norberg, P., Baugh, C. M., 
 Hawkins, E.,   
 Maddox, S., Madgwick, D., Lahav, O., Cole, S., Frenk, C. S., Baldry, I.,   
 Bland-Hawthorn, J., Bridges, T., Cannon, R., Colless, M., Collins, C.,   
 Couch, W., Dalton, G., De Propris, R., Driver, S. P., Efstathiou, G.,   
 Ellis, R. S., Glazebrook, K., Jackson, C., Lewis, I., Lumsden, S.,   
 Peacock, J. A., Peterson, B. A., Sutherland, W. \& Taylor, K.   
 2002 \mnras~ 332, 827   
   
 \bibitem[Osmer(1981)]{osmer81} Osmer, P. S. 1981 \apj~ 247, 762   
   
 \bibitem[Outram et al.(2003)]{outram} Outram, P. J., Hoyle, F.,  
 Shanks, T., Croom, S. M., Boyle, B. J., Miller, L., Smith, R. J. \&  
 Myers, A. D. 2003 \mnras~ 342 483 
  
 \bibitem[Peacock(1997)]{p97} Peacock, J. A. 1997 \mnras~ 284, 885   
   
 \bibitem[Peacock \& Dodds(1996)]{pd96}Peacock, J. A. \&  Dodds, S. J.    
 1996 \mnras~ 280, L19    
   
 \bibitem[Press \& Schechter(1974)]{ps74}Press, W. H.\& Schechter, P.    
 1974 \apj~ 187, 425    
   
 \bibitem[Reiprich \&  B\"ohringer(2002)]{rb02} Reiprich, T. H. \&    
 B\"ohringer, H. 2002 \apj~  567, 716   
   
 \bibitem[Schlegel et al.(1998)]{schlegel98} Schlegel, D. J., Finkbeiner,   
  D. P. \& Davis, M. 1998 \apj~ 500, 525   
   
 \bibitem[Seljak(2002)]{seljak02}Seljak, U. 2002 \mnras~ 337, 769   
   
 \bibitem[Shanks \& Boyle(1994)]{sb94} Shanks, T. \& Boyle, B. J. 1994   
 \mnras~ 271, 753   
   
 \bibitem[Shaver(1984)]{s84} Shaver, P. A., 1984, \aap~ 136, 9   
   
 \bibitem[Sheth \& Tormen(1999)]{st99} Sheth, R. K. \& Tormen, G.   
 1999 \mnras~ 308, 119   
   
 \bibitem[Smith et al.(2003)]{smith03}   
 Smith, R. E., Peacock, J. A., Jenkins, A., White, S. D. M.,    
 Frenk, C. S., Pearce F. R., Thomas, P. A., Efstathiou, G. \&    
 Couchman, H. M. P. 2003 \mnras~ 341, 1311   
   
 \bibitem[Steidel et al.(2002)]{shs02} Steidel, C. C., Hunt, M. P.,   
 Shapley, A. E., Adelberger, K. L., Pettini, M., Dickinson, M. \& Giavalisco,   
 M. 2002 \apj~ 576, 653 
   
 \bibitem[Tantalo \& Chiosi(2002)]{tantalo} Tantalo, R. \& Chiosi, C.   
 2002 \aap~ 388, 396   
   
 \bibitem[V\'eron-Cetty \& V\'eron(1984)]{VV84} V\'eron-Cetty, M. P. \&   
 V\'eron, P. 1984, Quasars and Active Galactic   
 Nuclei (5th Ed.); ESO Scientific Report   
   
 \bibitem[V\'eron-Cetty \& V\'eron(2001)]{VV01} V\'eron-Cetty, M. P. \&   
 V\'eron, P. 2001, Quasars and Active Galactic   
 Nuclei (10th Ed.); ESO Scientific Report   
   
 \bibitem[Viana et al.(2002)]{vnl02}Viana, P. T. P., Nichol, R. C. \&   
 Liddle, A. R. 2002, \apj~ 569, L75   
 
 \bibitem[Voges et al.(1999)]{vog99} Voges, W., Aschenbach, B., Boller, Th., 
 Br\"{a}uninger, H., Briel, U., Burkert, W., Dennerl, K., Englhauser, J., 
 Gruber, R., Haberl, F., Hartner, G., Hasinger, G., K\"urster, M., 
 Pfeffermann, E., Pietsch, W., Predehl, P., Rosso, C., Schmitt, J. H. M. M., 
 Tr\"umper, J. \& Zimmermann, H. U., 1999 \aap~ 349, 389 
 
 \bibitem[Woo \& Urry(2002)]{wu02} Woo, J. H. \& Urry, C. M. 2002   
 \apj~ 579,  530 
  
\end{thebibliography}
\end{document}